\documentclass[12pt]{iopart}
\usepackage{amssymb}
\usepackage[mathscr]{eucal}
\usepackage{epic}
\usepackage{eepic}
\newtheorem{prop}{Proposition}
\newtheorem{remark}{Remark}
\newtheorem{defin}{Definition}
\newtheorem{lem}{Lemma}

\newcommand{\nc}{\newcommand}
\nc{\A}{\mathcal{A}}                \nc{\R}{\mathbf{R}}
\nc{\ua}{u^{(1)}} \nc{\uz}{u^{(2)}} \nc{\ue}{u^{(3)}} \nc{\ug}{u^{(4)}}
\nc{\uh}{u^{(5)}} \nc{\wh}{w^{(5)}}
\nc{\wa}{w^{(1)}} \nc{\wz}{w^{(2)}} \nc{\we}{w^{(3)}} \nc{\wg}{w^{(4)}}
\nc{\uj}{u^{(8)}} \nc{\uk}{u^{(7)}} \nc{\ul}{u^{(6)}} \nc{\up}{u^{(5)}}
\nc{\wj}{w^{(8)}} \nc{\wk}{w^{(7)}} \nc{\wl}{w^{(6)}} \nc{\wo}{w^{(5)}}
\nc{\um}{u^{(\bar{5})}}  \nc{\wm}{w^{(\bar{5})}}
\nc{\dg}{\dagger} \nc{\dgg}{{\dagger\!\dagger}}
\nc{\dgh}{{\dagger\!\dagger\!\dagger}}
\nc{\ds}{\displaystyle} \nc{\XM}{\mathcal{X}}   \nc{\YM}{\mathcal{Y}}
\nc{\XB}{\Xi}      \nc{\YB}{\Upsilon}           \nc{\cc}{\mathfrak{C}}
\nc{\qop}{\mathbf{q}}   \nc{\pop}{\mathbf{p}}   \nc{\vop}{\mathbf{v}}
\nc{\uop}{\mathbf{u}}   \nc{\wop}{\mathbf{w}} 
\nc{\sop}{\mathbf{s}}   \nc{\xop}{\mathbf{x}}   \nc{\zop}{\mathbf{z}}
\nc{\Rop}{\mathcal{R}}  \nc{\Ropf}{\mathcal{R}^{(f)}}\nc{\Oc}{\mathcal{O}}
\nc{\Weyl}{\mathfrak{W}}\nc{\Eev}{\mathcal{E}}  \nc{\Eop}{\mathbf{E}}
\nc{\Uev}{\mathcal{U}}  \nc{\Uop}{\mathbf{U}}   
\nc{\Fop}{\mathbf{F}}   \nc{\Xop}{\mathbf{X}}
\nc{\one}{\mathbf{e}_1} \nc{\two}{\mathbf{e}_2} \nc{\thr}{\mathbf{e}_3}
\nc{\xh}{{\widehat{\xop}}}  \nc{\zh}{{\widehat{\zop}}} 
\nc{\uc}{\,{\sf u}}  \nc{\ub}{\bar{\sf u}} \nc{\xb}{\overline{x}}
\nc{\wc}{\:{\sf w}}  \nc{\wb}{\bar{\sf w}} \nc{\yb}{\overline{y}}
\nc{\vc}{\,{\sf v}}  \nc{\vb}{\bar{\sf v}}
\nc{\scc}{\,{\sf s}} \nc{\xc}{{\sf x}}     \nc{\yc}{{\:\sf y}}
\nc{\tc}{\,{\sf t}}  \nc{\zc}{{\sf z}}     \nc{\hc}{\:{\sf h}}
\nc{\beq}{\begin{equation}\fl}          \nc{\eeq}{\end{equation}}
\nc{\bea}{\begin{eqnarray}\fl}          \nc{\hx}{\hspace{3mm}}
\nc{\eea}{\end{eqnarray}}            \nc{\bdm}{\begin{displaymath}\fl}
\nc{\edm}{\end{displaymath}}         \nc{\hn}{\hspace*{-3mm}}
\nc{\ny}{\nonumber}        \nc{\om}{\omega}     \nc{\sig}{\sigma}
\nc{\hs}{\hspace{1cm}}     \nc{\hq}{\hspace{4mm}}
\nc{\ka}{\kappa}   \nc{\be}{\beta}     \nc{\al}{\alpha}
\nc{\si}{\sigma}
\nc{\ga}{\gamma}   \nc{\lam}{\lambda}  \nc{\Lam}{\Lambda}
\nc{\lk}{\left(}   \nc{\rk}{\right)}   \nc{\Rb}{\right]}
\nc{\lb}{\left\{}  \nc{\rb}{\right\}}  \nc{\Lb}{\left[}
\nc{\bl}{{\bar{\lambda}}}   \nc{\ra}{\rightarrow}
\nc{\xf}{\,\mathfrak{x}}  \nc{\yf}{\:\mathfrak{y}} \nc{\zf}{\,\mathfrak{z}}
\nc{\wf}{\,\mathfrak{w}}  \nc{\vf}{\:\mathfrak{v}} \nc{\pf}{\,\mathfrak{p}}
\nc{\qf}{\:\mathfrak{q}}  \nc{\mf}{\,\mathfrak{m}} \nc{\gf}{\,\mathfrak{g}}
\nc{\cf}{\,\mathfrak{c}}  \nc{\Nf}{\,\mathfrak{N}} \nc{\Zf}{\,\mathfrak{Z}}
\nc{\If}{\,\mathfrak{I}}  \nc{\Df}{\,\mathfrak{D}} \nc{\Uf}{\,\mathfrak{U}}
\nc{\Vf}{\,\mathfrak{V}}  \nc{\Af}{\,\mathfrak{A}} \nc{\BBf}{\,\mathfrak{B}}
\nc{\ff}{\,\mathfrak{f}}  \nc{\uf}{\,\mathfrak{u}} \nc{\ssf}{\,\mathfrak{a}}
\nc{\bbf}{\,\mathfrak{b}}\nc{\Thop}{\mathbf{\Theta}}\nc{\hf}{\,\mathfrak{h}}
\nc{\Rer}{\,\mathbb{R}}
\nc{\OT}{{\overline{\Theta}}}            \nc{\OTh}{{\overline{\Thop}}}
\nc{\OAf}{{\overline{\,\mathfrak{A}}}}   \nc{\OBf}{{\overline{\mathfrak{B}}}}
\nc{\hal}{{\textstyle\frac{1}{2}}}       \nc{\rN}{\right)^{\!N}\hn}
\nc{\rn}{\right)^{\!\frac{1}{N}}\!\!\!;\hn}   \renewcommand{\*}{\,\cdot\,}
\def\>{\rangle} \def\<{\langle}

\def\sk#1{\scriptstyle{#1}}\def\skk#1{\scriptscriptstyle{#1}}
\begin{document}

\title[Free parameterization of the MTE]{Explicit Free 
Parameterization of the Modified Tetrahedron Equation}
\author{G von Gehlen\dag,~ 
S Pakuliak\ddag~
 and S Sergeev\ddag\P}
\address{\dag\ Physikalisches Institut der Universit\"at Bonn, 
Nussallee 12, D-53115 Bonn, Germany}
\address{\ddag\ Bogoliubov Laboratory of Theoretical Physics,
Joint Institute for Nuclear Research, Dubna 141980, Moscow region, Russia}
\address{\P\  Max-Planck-Institut f\"ur Mathematik, Vivatsgasse 7,
D-53111 Bonn, Germany}
\ead{gehlen@th.physik.uni-bonn.de, pakuliak@thsun1.jinr.ru, 
sergeev@mpim-bonn.mpg.de, sergeev@thsun1.jinr.ru}

\begin{abstract}
The Modified Tetrahedron Equation (MTE) with affine Weyl quantum variables at 
$N$-$th$ root of unity is solved by a rational mapping operator which is 
obtained from the solution of a linear problem. We show that the solutions 
can be 
parameterized in terms of eight free parameters and sixteen discrete phase 
choices, thus providing a broad starting point for the construction of 
3-dimensional 
integrable lattice models. The Fermat curve points parameterizing the 
representation of the mapping operator in terms of cyclic functions are 
expressed in terms of the independent parameters. An explicit formula for the 
density factor of the MTE is derived. For the example $N=2$ we write the MTE 
in full detail. We also discuss a solution of the MTE in terms of bosonic 
continuum functions.
\end{abstract}
\submitto{\JPA}
\pacs{05.45-a,05.50+q}
\section*{Introduction}
The Zamolodchikov tetrahedron equation is the condition \cite{Zamo} for 
the existence of a commuting set of layer-to-layer transfer matrices for 
3-dimensional lattice
models, in much the same way as the Yang-Baxter equation is the analogous
condition in the 2-dimensional case. Only very few solutions to these very
restrictive equations have been found \cite{Zamo,Bazh-Bax}.
So various Modified Tetrahedron Equations (MTE)
have been studied to which more solutions can be obtained
\cite{bms-mte,sms-vertex,s-old,ms-mte, s-kirhgoff,s-symplectic}, still leading to
commuting transfer matrices or generating functionals for conserved quantities.

In this paper we shall concentrate on a particular 
MTE proposed in \cite{s-qem,s-kiev}. The quantum variables are elements 
from an ultra-local affine Weyl algebra attached to every vertex of a 
2-dimensional graph. Since we consider the 
Weyl parameter to be a $N$-$th$ root of unity, the $N$-$th$ powers
of the quantum variables form a classical system which determines the 
parameters of the quantum system, as has been considered in for the
discrete sin-Gordon model and other models recently, e.g. 
\cite{Fa-Vo,fk-qd,br-qd,br-unpublished}. 
So the parameters of the eight $\:\R\:$ matrices appearing in the MTE are 
different, but related by functional mappings. 

A linear problem discussed previously by one of us 
\cite{s-qem, s-kiev}, is used to determine the mapping which provides a 
multi-parameter solution to the MTE. The construction of a generating 
functional of the conserved quantities has been given in \cite{s-qem}. 
Here we concentrate on calculating the density factor of the MTE and to 
give a useful choice of the eight continuous parameters of the mapping.

The aim of studying these equations is at least two-fold: first a
2+1-dimensional integrable lattice model should emerge, and second, the MTE
can be used by contraction \cite{s-rma} to construct new 2-dimensional 
lattice models with parameters living on higher Riemann surfaces.

The paper is organized as follows: In Sec.1 we introduce the rational 
mapping in the affine Weyl space and show that if the Weyl parameter is 
a root of unity this splits into a matrix mapping and a functional mapping. 
Then we consider two different realizations: in terms of cyclic functions 
(this will be mainly used) and in terms of Gaussians. 
Sec. 2 discusses the modified tetrahedron equation and we calculate its 
weight function. In Sec. 3 we focus on the parameterization in terms of 
line ratios, finding eight continuous parameters and analyze the phase 
ambiguities. For the specific case $N=2$ in Sec. 4 we show that the
modified tetrahedron equations can be written quite explicitly and that 
of their $2^{12}$ matrix components there are 256 linearly independent 
equations. In Sec. 5 we give a summary and mention future applications.

\section{The rational mapping $\Rop$ in the space of a triple affine Weyl
algebra.}

The central object of our considerations will be a mapping operator acting
in the space of a triple Weyl affine algebra. We shall see that this 
mapping operator can be written as a superposition of a functional mapping and
a finite-dimensional similarity transformation. It is the operator of this
similarity transformation which will satisfy the MTE. Several interpretations 
are possible, e.g. as vertex Boltzmann weights (albeit not positive ones) of a 
three-dimensional lattice model. It will be a generalization of the 
Zamolodchikov-Bazhanov-Baxter \cite{Zamo,Bazh-Bax} Boltzmann weights in the 
Sergeev-Mangazeev-Stroganov \cite{sms-vertex} vertex formulation.
The principle from which this mapping is obtained has been described in 
detail in \cite{s-qem}. It is a current conservation principle with a 
Baxter Z-invariance. 

\subsection{The linear problem.}\label{Lin-pro}
\begin{figure}
\setlength{\unitlength}{0.0005in}
\begin{center}
{\renewcommand{\dashlinestretch}{30}
\begin{picture}(7909,5664)(0,-10)
\path(4297,5637)(7897,5637)(6322,4062)
    (2722,4062)(4297,5637)
\path(4297,1587)(7897,1587)(6322,12)
    (2722,12)(4297,1587)
\path(4297,5187)(6997,5187)
\blacken\path(6877.000,5157.000)(6997.000,5187.000)
(6877.000,5217.000)(6877.000,5157.000) \path(3847,4287)(4972,5412)
\blacken\path(4908.360,5305.934)(4972.000,5412.000)
(4865.934,5348.360)(4908.360,5305.934) \path(3622,4512)(6772,5412)
\blacken\path(6664.859,5350.188)(6772.000,5412.000)
(6648.375,5407.879)(6664.859,5350.188) \path(3847,462)(6547,462)
\blacken\path(6427.000,432.000)(6547.000,462.000)
(6427.000,492.000)(6427.000,432.000) \path(5872,237)(6997,1362)
\blacken\path(6933.360,1255.934)(6997.000,1362.000)
(6890.934,1298.360)(6933.360,1255.934) \path(3622,237)(7447,1317)
\blacken\path(7339.667,1255.521)(7447.000,1317.000)
(7323.363,1313.264)(7339.667,1255.521)
\dashline{60.000}(5602,4062)(6007,5187)
\blacken\path(5994.580,5063.932)(6007.000,5187.000)
(5938.127,5084.255)(5994.580,5063.932)
\dashline{60.000}(5062,4062)(4747,5187)
\blacken\path(4808.245,5079.533)(4747.000,5187.000)
(4750.467,5063.355)(4808.245,5079.533)
\dashline{60.000}(4702,4062)(4252,4692)
\blacken\path(4346.161,4611.789)(4252.000,4692.000)
(4297.337,4576.915)(4346.161,4611.789) \path(6097,462)(5062,4062)
\path(4432,462)(5602,4062) \path(6760,1120)(4702,4062)
\path(1597,2037)(1597,1587)(2497,1587)
    (2497,2037)(1597,2037)(1417,2217)
\blacken\path(1523.066,2153.360)(1417.000,2217.000)
(1480.640,2110.934)(1523.066,2153.360) \thicklines
\path(22,2262)(2722,2262)
\blacken\path(2482.000,2202.000)(2722.000,2262.000)
(2482.000,2322.000)(2482.000,2202.000) \path(1372,912)(1372,3612)
\blacken\path(1432.000,3372.000)(1372.000,3612.000)
(1312.000,3372.000)(1432.000,3372.000) \thinlines
\dashline{60.000}(4072,687)(2227,1137)
\blacken\path(2350.691,1137.711)(2227.000,1137.000)
(2336.474,1079.420)(2350.691,1137.711) \path(4297,732)
(4237.119,714.392)
    (4200.020,704.083)
    (4160.950,692.921)
    (4122.012,681.019)
    (4085.306,668.490)
    (4027.000,642.000)
\path(4027,642) (3992.474,616.589)
    (3953.222,581.242)
    (3917.110,542.525)
    (3892.000,507.000)
\path(3892,507) (3869.200,460.073)
    (3847.866,401.258)
    (3837.349,339.063)
    (3847.000,282.000)
\path(3847,282) (3887.714,229.806)
    (3948.190,190.635)
    (4014.321,163.397)
    (4072.000,147.000)
\path(4072,147) (4128.683,138.773)
    (4195.659,136.761)
    (4231.881,137.733)
    (4269.326,139.836)
    (4307.544,142.928)
    (4346.084,146.869)
    (4384.496,151.517)
    (4422.331,156.730)
    (4459.138,162.369)
    (4494.467,168.292)
    (4558.890,180.425)
    (4612.000,192.000)
\path(4612,192) (4663.518,207.673)
    (4726.761,231.529)
    (4788.874,258.120)
    (4837.000,282.000)
\path(4837,282) (4879.455,308.020)
    (4930.997,343.406)
    (4980.542,381.840)
    (5017.000,417.000)
\path(5017,417) (5045.042,452.197)
    (5075.826,497.756)
    (5099.697,547.937)
    (5107.000,597.000)
\path(5107,597) (5091.355,650.383)
    (5058.726,702.259)
    (5016.484,746.505)
    (4972.000,777.000)
\path(4972,777) (4909.336,791.563)
    (4872.449,792.170)
    (4834.176,789.945)
    (4796.257,786.102)
    (4760.428,781.854)
    (4702.000,777.000)
\path(4702,777) (4642.667,778.429)
    (4606.048,779.679)
    (4567.412,780.810)
    (4528.725,781.465)
    (4491.951,781.286)
    (4432.000,777.000)
\path(4432,777) (4363.656,756.337)
    (4297.000,732.000)
\path(1372,912) (1442.930,919.968)
    (1508.875,927.767)
    (1570.080,935.448)
    (1626.794,943.060)
    (1679.264,950.652)
    (1727.736,958.274)
    (1772.459,965.974)
    (1813.679,973.804)
    (1851.643,981.811)
    (1886.599,990.046)
    (1948.474,1007.396)
    (2001.282,1026.249)
    (2047.000,1047.000)
\path(2047,1047)    (2081.547,1065.839)
    (2118.788,1088.434)
    (2158.205,1114.405)
    (2199.276,1143.375)
    (2241.482,1174.963)
    (2284.304,1208.792)
    (2327.221,1244.482)
    (2369.714,1281.656)
    (2411.262,1319.935)
    (2451.347,1358.939)
    (2489.448,1398.290)
    (2525.045,1437.610)
    (2557.618,1476.519)
    (2586.649,1514.640)
    (2611.616,1551.593)
    (2632.000,1587.000)
\path(2632,1587)    (2659.338,1654.703)
    (2670.663,1694.173)
    (2680.551,1736.602)
    (2689.099,1781.402)
    (2696.402,1827.982)
    (2702.556,1875.753)
    (2707.656,1924.125)
    (2711.799,1972.509)
    (2715.079,2020.315)
    (2717.592,2066.953)
    (2719.434,2111.835)
    (2720.700,2154.370)
    (2721.486,2193.969)
    (2722.000,2262.000)
\path(2722,2262)    (2721.311,2330.056)
    (2720.335,2369.682)
    (2718.835,2412.251)
    (2716.734,2457.169)
    (2713.955,2503.846)
    (2710.423,2551.688)
    (2706.059,2600.104)
    (2700.787,2648.500)
    (2694.530,2696.286)
    (2687.211,2742.868)
    (2678.754,2787.655)
    (2669.081,2830.054)
    (2658.115,2869.472)
    (2632.000,2937.000)
\path(2632,2937)    (2612.009,2973.931)
    (2587.649,3012.973)
    (2559.381,3053.663)
    (2527.669,3095.539)
    (2492.976,3138.138)
    (2455.764,3180.996)
    (2416.498,3223.650)
    (2375.639,3265.639)
    (2333.650,3306.498)
    (2290.996,3345.764)
    (2248.138,3382.976)
    (2205.539,3417.669)
    (2163.663,3449.381)
    (2122.973,3477.649)
    (2083.931,3502.009)
    (2047.000,3522.000)
\path(2047,3522)    (1979.472,3548.115)
    (1940.054,3559.081)
    (1897.655,3568.754)
    (1852.868,3577.211)
    (1806.286,3584.530)
    (1758.500,3590.787)
    (1710.104,3596.059)
    (1661.688,3600.423)
    (1613.846,3603.955)
    (1567.169,3606.734)
    (1522.251,3608.835)
    (1479.682,3610.335)
    (1440.056,3611.311)
    (1372.000,3612.000)
\path(1372,3612)    (1303.944,3611.311)
    (1264.318,3610.335)
    (1221.749,3608.835)
    (1176.831,3606.734)
    (1130.154,3603.955)
    (1082.312,3600.423)
    (1033.896,3596.059)
    (985.500,3590.787)
    (937.714,3584.530)
    (891.132,3577.211)
    (846.345,3568.754)
    (803.946,3559.081)
    (764.528,3548.115)
    (697.000,3522.000)
\path(697,3522) (660.071,3502.008)
    (621.030,3477.646)
    (580.340,3449.377)
    (538.465,3417.665)
    (495.867,3382.971)
    (453.009,3345.760)
    (410.354,3306.494)
    (368.365,3265.635)
    (327.506,3223.647)
    (288.238,3180.993)
    (251.026,3138.136)
    (216.333,3095.538)
    (184.620,3053.662)
    (156.352,3012.972)
    (131.991,2973.931)
    (112.000,2937.000)
\path(112,2937) (85.885,2869.472)
    (74.919,2830.054)
    (65.246,2787.655)
    (56.789,2742.868)
    (49.470,2696.286)
    (43.213,2648.500)
    (37.941,2600.104)
    (33.577,2551.688)
    (30.045,2503.846)
    (27.266,2457.169)
    (25.165,2412.251)
    (23.665,2369.682)
    (22.689,2330.056)
    (22.000,2262.000)
\path(22,2262)  (22.689,2193.944)
    (23.665,2154.318)
    (25.165,2111.749)
    (27.266,2066.831)
    (30.045,2020.154)
    (33.577,1972.312)
    (37.941,1923.896)
    (43.213,1875.500)
    (49.470,1827.714)
    (56.789,1781.132)
    (65.246,1736.345)
    (74.919,1693.946)
    (85.885,1654.528)
    (112.000,1587.000)
\path(112,1587) (131.992,1550.071)
    (156.354,1511.030)
    (184.623,1470.340)
    (216.335,1428.465)
    (251.029,1385.867)
    (288.240,1343.009)
    (327.506,1300.354)
    (368.365,1258.365)
    (410.353,1217.506)
    (453.007,1178.238)
    (495.864,1141.026)
    (538.462,1106.333)
    (580.338,1074.620)
    (621.028,1046.352)
    (660.069,1021.991)
    (697.000,1002.000)
\path(697,1002) (743.831,981.831)
    (797.433,964.737)
    (859.785,950.456)
    (894.860,944.287)
    (932.864,938.723)
    (974.044,933.729)
    (1018.647,929.274)
    (1066.921,925.324)
    (1119.113,921.847)
    (1175.469,918.808)
    (1236.238,916.176)
    (1301.666,913.918)
    (1372.000,912.000)
\put(1310,2190){\makebox(0,0)[lb]{$\bullet$}}
\put(1560,1550){\makebox(0,0)[lb]{$
\begin{array}{c}\vspace{-2mm}\!
\mathbf{u}_3\;\mathbf{w}_3\\[-2mm]
\:\kappa_3
\end{array}
$}} \put(652,2757){\makebox(0,0)[lb]{$\phi_a$}}
\put(1912,2802){\makebox(0,0)[lb]{$\phi_b$}}
\put(607,1677){\makebox(0,0)[lb]{$\phi_c$}}
\put(1732,1227){\makebox(0,0)[lb]{$\phi_d$}}
\put(4342,220){\makebox(0,0)[lb]{$\mathfrak{w}_3$}}
\put(6097,192){\makebox(0,0)[lb]{$\mathfrak{w}_1$}}
\put(6817,867){\makebox(0,0)[lb]{$\mathfrak{w}_2$}}
\put(5250,3117){\makebox(0,0)[lb]{$\bullet$}}
\put(3937,4737){\makebox(0,0)[lb]{$\mathfrak{w}'_2 $}}
\put(4400,5277){\makebox(0,0)[lb]{$\mathfrak{w}'_1$}}
\put(5827,5232){\makebox(0,0)[lb]{$\mathfrak{w}'_3$}}
\put(4402,3027){\makebox(0,0)[lb]{$\mathcal{R}_{1,2,3}$}}
\end{picture}
}
\caption{\footnotesize{The linear problem for the  vertex with 
associated Weyl pair $\mathbf{u}_3$, $\mathbf{w}_3$ 
and parameter $\kappa_3$ and the visualization of $\mathcal{R}_{1,2,3}$. 
The elements $\mathfrak{w}_1$, $\mathfrak{w}'_1$, etc. of the 
ultra-local affine Weyl 
algebras are assigned to the vertices of two auxiliary two dimensional 
lattices formed by the intersection of three straight lines with the 
auxiliarly planes.}} \label{vertex}
\end{center}
\end{figure}
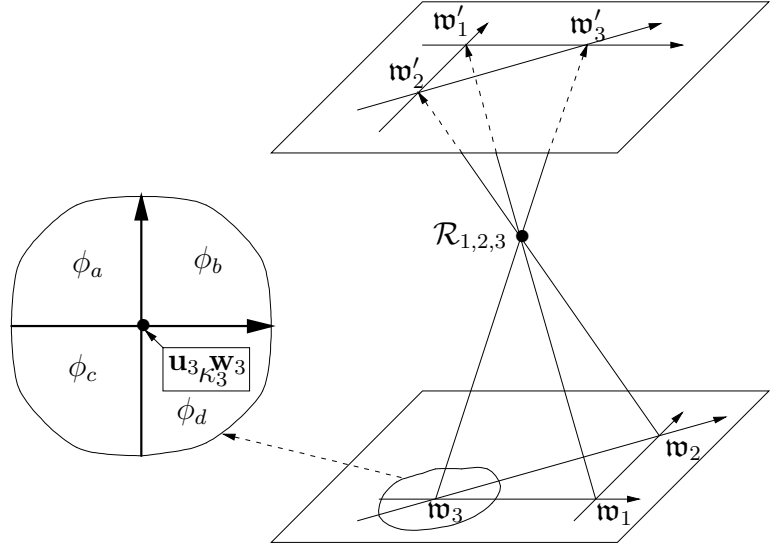

To set the framework we assign to each vertex $j$ of a $2d$ graph the elements 
$\uop_j,\;\wop_j$ of an affine Weyl algebra at Weyl parameter $q$ a root 
of unity:
\beq \label{Weylu}  \uop_j\*\wop_j\;=\;q\;\wop_j\*\uop_j,\hs q\;=\;\om\;
\stackrel{\mathrm{def}}{=}\;
e^{2\pi\,i/N}\,,\hs N\in\mathbb{Z}\;,\;\;
N\;\ge\;2\;. \end{equation}
Since $q$ is a root of unity, $\uop_j^N$ and $\wop_j^N$ are centers of the 
Weyl algebra. We shall often represent the canonical pair 
$\:(\uop_j,\:\wop_j)\;$ by its action on a cyclic basis as unitary $N\times N$ 
matrices multiplied by complex parameters $\;u_j,\:w_j,$ writing 
\beq\label{uw-xz}
\ds \uop\;=\;u\,\xop\;;\hs\hs\wop\;=\;w\,\zop\;;
\end{equation}
\beq |\,\sigma\rangle\;\equiv\;|\,\sigma\;\;mod\;N\>\;;\hq
\langle\,\sigma\,|\,\sigma' \rangle\;=\;\delta_{\sigma,\sigma'};\hq
\xop\,|\,\sigma\rangle\;=\;|\,\sigma\rangle\,\om^\sigma\;;\hq
\zop\,|\,\sigma\rangle\;=\;|\,\sigma+1\rangle\:. \label{XZdef}  \end{equation}
The centers are represented by numbers:
\beq   \uop_j^N\;=\;u_j^N\;,\hs\hs  \wop_j^N\;=\;w_j^N\;.
\end{equation}
We define the ultra-local Weyl algebra $\Weyl^{\otimes\Delta}$
 as the tensor product of $\Delta$ copies of Weyl pairs
\beq
\ds \uop_j\;=\; 1 \otimes 1 \otimes ... \,\otimes
\underbrace{\uop}_{{j-th\atop place}}\otimes\, ...\;\:; \hq
\wop_j\;=\; 1 \otimes 1 \otimes ...\, \otimes
\underbrace{\wop}_{{j-th\atop place}}\otimes\, ...\;.
\end{equation} 
Denote the four faces around the vertex $j$ (in which two oriented 
lines cross) 
clockwise by $a,\;b,\;c,\;d\;$ as e.g. shown in the left side of 
Fig.~\ref{vertex} for the vertex $j=3$. 
Imagine a current $\langle\,\phi\,|$ flowing out of 
the vertex into the four faces $a,\;b,\;c,\;d$, each face receiving a 
current $\;\langle\,\phi_s|\;\;\;(s=a,b,c,d)\;$ according to the values of the 
Weyl variables sitting at the vertex and a coupling constant $\kappa_j\:$ 
(which may be different at each vertex):
\beq \langle\,\phi\,|\;=\;\langle\phi_a|\;+\,\langle\phi_b|\cdot q^{1/2}
\uop_j\;+\,\langle\phi_c|\,\cdot\wop_j\;+\,\langle\phi_d|\cdot\kappa_j\;
\uop_j\;\wop_j.  \label{lin-sys}\end{equation}
Demanding that the total current flowing out of an internal vertex is zero: 
$\langle\,\phi\,|\,=\,0\,,$ and demanding also 
that the currents flowing into the outer faces of various graphs are 
independent of the internal structure of the graphs (this is a Z-invariance 
assumption),
we get a condition for the equivalence of linear problems. The right hand 
side of Fig.1 shows two such linear problems, one in the bottom plane,
another in the upper plane. The equivalence condition determines the  
mapping  $\:\Rop_{1,2,3}\;$ between the lower 
$({\mathfrak{w}_1},\,{\mathfrak{w}_2},\,{\mathfrak{w}_3})$ and upper
$({\mathfrak{w}'_1},\,{\mathfrak{w}'_2},\,{\mathfrak{w}'_3})$
triangle in Fig.~\ref{vertex} uniquely. 
The details of this calculation can be found in \cite{s-qem}, 
in (\ref{rma})-(\ref{rmc}) below we present the result.\\
For our case of interest $\;q=\om\:$ it is convenient to choose the specific 
form (\ref{lin-sys}) of the coefficients, which is unsymmetrical in 
$a,\;b,\;c,\;d$. There exists a fully symmetrical formulation of the linear 
problem valid at general $q$, still leading to a unique mapping 
$\:\Rop_{1,2,3}$ \cite{s-qem}. However, this will not be needed here.

\subsection{The rational mapping $\Rop_{1,2,3}$}
\label{R-mapping}

The solution of the equivalence problem of the linear current flows is the 
following rational mapping $\;\Rop\;$ acting in the ring of rational functions 
of the generators of the ultra-local Weyl algebra \cite{s-qem}:  
For any rational function $\Phi$ we define
\beq\label{r-mapping}
\ds\lk\Rop_{1,2,3}\circ\Phi\rk
(\uop_1^{},\wop_1^{},\uop_2^{},\wop_2^{},
\uop_3^{},\wop_3^{},...)\stackrel{def}{=}
\Phi(\uop_1',\wop_1',\uop_2',\wop_2',\uop_3',
\wop_3',...) \label{rma}\end{equation}
where in the right hand side of (\ref{r-mapping}) the 
$\uop_\al$ and $\wop_\al$ 
remain unchanged for all $\al\not\in\{1,2,3\}$, and the primed elements 
are rational functions of $\uop_1,...,\wop_3$, given by the definition
\beq\label{mapping}
\ds\left.\begin{array}{lll}
\ds\wop_1'\;=\;\ds\wop_2^{}\*\Lam_{3}^{}\;, &
\ds\wop_2'\;=\;\ds\Lam_{3}^{-1}\* \wop_1^{}\;,
&
\ds\wop_3'\;=\;\ds\Lam_{2}^{-1}\* \uop_1^{-1}\;, \\&&\\[-2mm]
\ds\uop_1'\;=\;\ds\Lam_{2}^{-1}\* \wop_3^{-1}\;,&
\ds\uop_2'\;=\;\ds\Lam_{1}^{-1}\* \uop_3^{}\;,&
\ds\uop_3'\;=\;\ds\uop_2^{}\*\Lam_{1}^{}\;,\\
\end{array}\right.\label{rmb}\end{equation}
where
\beq\label{lambda}\ds\left.\begin{array}{ccl}
\ds\Lam_{1}^{}
& \equiv & \ds
\uop_1^{-1}\*\uop_3^{}\,-\,
q^{1/2}\,\uop_1^{-1}\*\wop_1^{} \,+\,
\ka_1^{}\;\wop_1^{}\*\uop_2^{-1}\;,\\[-2mm]
&&\\
\ds\Lam_{2}^{} & \equiv & \ds
{\ka_1\over\ka_2}\;\uop_2^{-1}\*\wop_3^{-1}\,+\,
{\ka_3\over\ka_2}\;\uop_1^{-1}\*\wop_2^{-1}\,-\,
q^{-1/2}\;{\ka_1\;\ka_3\over\ka_2}\;
\uop_2^{-1}\*\wop_2^{-1}\;,\\[-2mm]
&&\\
\ds\Lam_{3}^{} & \equiv & \ds \wop_1^{}\*\wop_3^{-1}\,-\,
q^{1/2}\;\uop_3^{}\*\wop_3^{-1}\,+\,
\ka_3^{}\;\wop_2^{-1}\*\uop_3\;.\\
\end{array}\right.\label{rmc}\end{equation}
$\ka_{1},\ka_2,\ka_3\;\in\;\mathbb{C}$ are arbitrary extra parameters of the 
mapping $\Rop_{1,2,3}$. In this subsection $q$ can be in general position.

Note that the order of the factors in the right hand sides of
(\ref{mapping}) does not matter. Each combination $\Lambda_i$
$\:(i=1,2,3)\:$ contains only elements of the Weyl algebra
which commute with the other factors in the product.
For example, $\Lambda_3$ does not have the Weyl operators $\uop_1$
and $\uop_2$. From (\ref{mapping}) we see that the rational mapping 
$\Rop_{1,2,3}$ has the three invariants:
\beq\label{centers}  \wop_1\wop_2,\hs\uop_2\uop_3,\hs\uop_1\wop_3^{-1}.
\end{equation}
This means that
this mapping has the property that the products ${\uop_j}^{-1}{\uop_j}'$
and ${\wop_j}^{-1}{\wop_j}'$ for $j=1,\,2,\,3$ (no summation over $j$)
depend only on three operators which we denote by $\uop,\;\vop$ and $\wop$:
\beq
\uop=\wop_2^{-1}\wop_3;\hs\vop=\uop_1\uop_2^{-1};
\hs\wop=\wop_1\uop_3^{-1},        \label{indop}\end{equation}
as one easily checks explicitly:
\bea 
 (\wop_3^{-1}{\wop_3}')^{-1}& =& (\uop_1^{-1}{\uop_1}')^{-1}=
 \uop_1\Lam_2\wop_3\;=\;  \frac{\ka_1}{\ka_2}\vop+\frac{\ka_3}{\ka_2}
  \uop-q^{-1/2}\frac{\ka_1\ka_3}{\ka_2}\vop\uop;\ny\\
\fl (\uop_2^{-1}{\uop_2}')^{-1}& =&  {\uop_3}'\uop_3^{-1}\;\;\;\;=\;
  \uop_3^{-1}\Lam_1\uop_2\;=\;\vop^{-1}+\ka_1\wop\;
  -\:q^{1/2}\vop^{-1}\wop;\ny\\[2mm]
\fl \wop_1^{-1}{\wop_1}'\hx\hx & =&
(\wop_2^{-1}{\wop_2}')^{-1}=
   \wop_2\Lam_3\wop_1^{-1}\;=\;\uop^{-1}+\ka_1\wop^{-1}
   -q^{1/2}\wop^{-1}\uop^{-1}.\eea
Observe that the three operators $\;$(\ref{indop})$\;$ form a triple Weyl
algebra:
$\;\vop\uop\,=\,q\,\uop\vop,\;\;\;\vop\wop\,=\,q\,\wop\vop,\;\;\;
\uop\wop\,=\,q\,\wop\uop,$ and also at each vertex $j$ we can regard
$\;\uop_j,\:\wop_j$ together with $\vop_j\equiv\ka_j\uop_j\wop_j$ as forming
triple Weyl algebras.

The mapping $\Rop$ has the property (see
\cite{s-kirhgoff,s-old,s-tmf1,s-qem,s-kiev} for the details):
\begin{prop}
The invertible mapping $\Rop_{i,j,k}$ is an automorphism of
$\Weyl^{\otimes\Delta}$.
\end{prop}
\begin{remark}
The Proposition states that the rational mapping (\ref{mapping})
is canonical, namely, it sends three copies of the ultra-local
Weyl algebras into the same Weyl algebras. \end{remark}

\noindent
Later, in Sec. \ref{moteeq}, Proposition \ref{tet-prop}, we shall 
discuss the second crucial property
of the mapping $\Rop$: it solves the Tetrahedron Equation (\ref{op-te}).

\subsection{Functional part at root of unity}\label{fuN}

In all the following we shall consider only the case that $q$ is a root 
of unity (\ref{Weylu}) and use the unitary representation (\ref{XZdef}) 
of $\;\Weyl^{\otimes\Delta}\,.\;$
In this representation each affine Weyl element $\uop_j$ and $\wop_j$ 
will contain one free parameter $u_j$ resp.
$w_j$, as written in (\ref{uw-xz}).

The basic fact is that at Weyl parameter root of unity any rational
automorphism of the ultra-local Weyl algebra implies a rational
mapping in the space of the $N$-th powers of the parameters of the
representation \cite{fk-qd,br-qd,br-unpublished}. In our case 
(\ref{mapping}) it is easy to check that the mapping
$\,\Rop_{1,2,3}$ implies
\beq\label{mapping-N}
\ds\left.\begin{array}{lll}
\ds \wop_1^{\prime N}\;=\;\ds w_2^{N}\, \Lam_3^{N}\;, &
\ds \wop_2^{\prime N}\;=\;\ds {w_1^{N}\over\Lam_3^{N}}\;, &
\ds \wop_3^{\prime N}\;=\;\ds {1\over\Lam_2^{N}\, u_1^{N}}\;, \\[5mm]
\ds \uop_1^{\prime N}\;=\;\ds {1\over \Lam_2^{N}\, w_3^{N}}\;,&
\ds \uop_2^{\prime N}\;=\;\ds {u_3^N\over\Lam_1^{N}}\;,&
\ds \uop_3^{\prime N}\;=\;\ds u_2^{N}\, \Lam_1^{N}\;,
\end{array}\right.
\end{equation}
where the $N$-th powers of the $\Lam_k$ are also numbers:
\\[1mm]
\beq\label{lambda-N}
\ds\left.\begin{array}{ccl}
\ds\Lam_1^N & = & \ds
u_1^{-N}\,u_3^{N}\;+\;u_1^{-N}\,w_1^{N}\;+\;
                           \ka_1^N\;w_1^{N}\,u_2^{-N}\;,\\
&&\\
\ds\Lam_2^N & = & \ds
{\ka_1^N\over\ka_2^N}\;u_2^{-N}\, w_3^{-N}\;+\;
{\ka_3^N\over\ka_2^N}\;u_1^{-N}\, w_2^{-N}\;+\;
{\ka_1^N\;\ka_3^N\over\ka_2^N}\;
u_2^{-N}\,w_2^{-N}\;,\\
&&\\
\ds\Lam_3^N & = & \ds
w_1^{N}\, w_3^{-N}\;+\;
u_3^{N}\, w_3^{-N}\;+\;
\ka_3^N\;w_2^{-N}\,u_3^{N}\;
\end{array}\right.
\end{equation}
since for $\;q=\om\:$ and $\;a,b\in {\mathbb{C}}\;$
one has $\;\lk\,a\,\uop\,+\,b\,\wop \rk^N\,=\,(a\,u)^N \,+\,(b\,w)^N$,
using $\;\sum_{j=\mathbb{Z}_N}\om^j=0$.

\begin{defin}
The functional counterpart of the mapping $\Rop_{1,2,3}$
is the mapping $\Rop_{1,2,3}^{(f)}$, acting on the space
of functions of the parameters $u_j,\:w_j\;\:(j=1,2,3)$
\beq\ds \lk\Rop_{1,2,3}^{(f)}\circ\phi\rk
(u_1^{},w_1^{},u_2^{},w_2^{},u_3^{},w_3^{})\;
\stackrel{def}{=}\;\phi(u_1',w_1',u_2',w_2',u_3',w_3')\;,
   \label{fctm} \end{equation}
where the primed variables are functions of the unprimed ones,
defined via
\beq \ds u_1^{\prime N}\,=\,\uop_1^{\prime N}\;,\;\;\;
w_1^{\prime N}\,=\,\wop_1^{\prime N}\;,\;\;\;
\mbox{etc.,}\end{equation}
such that the $\;u_j,\:w_j,\;u_j',\;w_j'$ satisfy
\beq\label{f-centers}\ds w_1'w_2'\,=\,w_1^{}w_2^{}\;,\;\;\;
u_2'u_3'\,=\,u_2u_3\;,\;\;\;
{u_1'\over w_3'}\,=\,{u_1^{}\over w_3^{}}\;.  \end{equation}
The three free phases of the $N$-th roots are extra discrete parameters of
$\:\Rop_{1,2,3}^{(f)}$.
\end{defin}
We use the invariance of the three centers $\:u_2u_3,\;u_1/w_3$ and
$\:w_1w_2$ to define three functions $\;\Gamma_1,\;\Gamma_2,\;\Gamma_3$:
\bea
\Gamma_1^N&=& \frac{{u_3'}^N}{u_3^N}=\frac{u_2^N}{{u_2'}^N}=
\lk u_3^{-1}\Lambda_1 u_2\rk^N\;=\;\frac{u_2^N}{u_1^N}+
    \frac{w_1^Nu_2^N}{u_1^Nu_3^N}
   +\ka_1^N\frac{w_1^N}{u_3^N},\ny \\
\fl \Gamma_2^N&=& \frac{u_1^N}{{u_1'}^N}=\frac{w_3^N}{{w_3'}^N}=
\lk w_3 \Lambda_2 u_1\rk^N\;=\;\frac{\ka_1^N u_1^N}{\ka_2^N u_2^N}
 +\!\frac{\ka_3^N w_3^N}{\ka_2^N w_2^N}+\!\frac{\ka_1^N
\ka_3^N u_1^N w_3^N}{\ka_2^N u_2^N w_2^N};\ny\\
\fl \Gamma_3^N&=& \frac{{w_1'}^N}{w_1^N}=\frac{w_2^N}{{w_2'}^N}=
\lk w_1^{-1}\Lambda_3 w_2\rk^N\;=\;\frac{w_2^N}{w_3^N}+
    \frac{w_2^N u_3^N}{w_1^N w_3^N}+\ka_3^N\frac{u_3^N}{w_1^N}.\eea
so that, alternatively to using (\ref{mapping-N}),$\;$(\ref{lambda-N}),
the functional mapping can be written as
\[\fl \Rop^{(f)}\circ w_1=w_1\Gamma_3, \hq
   \Rop^{(f)}\circ w_2=\frac{w_2}{\Gamma_3}, \hq
   \Rop^{(f)}\circ w_3=\frac{w_3}{\Gamma_2}, \]
\[\fl \Rop^{(f)}\circ u_1=\frac{u_1}{\Gamma_2}, \hq
   \Rop^{(f)}\circ u_2=\frac{u_2}{\Gamma_1}, \hq
   \Rop^{(f)}\circ u_3=u_3\Gamma_1. \]
The $\Gamma_j$ depend on 3 variables and the 3 constants $\ka_j$. 
Their phases are arbitrary.

\subsection{Matrix part at root of unity}

Now we consider the matrix structure of $\Rop_{1,2,3}$
at $q$ a root of unity. First of all, we define
\beq\label{matrixes}
\ds\xop_1'\;=\;{\uop_1'\over u_1'}\;,\;\;\;
\zop_1'\;=\;{\wop_1'\over w_1'}\;,\;\;\;
\mbox{etc.}
\end{equation}
The normalization implies the conservation of the centers
\beq
\ds\xop_1^{\prime N}\,=\,\xop_1^N\,=\,1\;,\;\;\;
\zop_1^{\prime N}\,=\,\zop_1^N\,=\,1\;,\;\;\;
\mbox{etc.}
\end{equation}
The rational mapping (\ref{mapping}) for the set of matrices
(\ref{matrixes}) has the form
\bea\label{xzprime}
\lk\xop_1'\rk^{-1}&=&\frac{\ka_1u'_1}{\ka_2u_2}\xop_2^{-1}
  +\frac{\ka_3u'_1w_3}{\ka_2u_1w_2}\xop_1^{-1}\zop_2^{-1}\zop_3
  -\om^{1/2}\frac{\ka_1\ka_3u'_1w_3}
  {\ka_2u_2w_2}\xop_2^{-1}\zop_2^{-1}\zop_3, \ny\\[2mm]
\fl \zop_1'&=&\frac{w_2w_1}{w'_1w_3} \zop_1 \zop_2 \zop_3^{-1}
  -\om^{1/2}\frac{w_2u_3}{w'_1w_3}\zop_2 \xop_3 \zop_3^{-1}
  +\frac{\ka_3}{w'_1}\xop_3, \ny \\[2mm]
\fl \lk\xop_2'\rk^{-1}&=&\frac{u'_2}{u_1}\xop_1^{-1}-\om^{1/2}
\frac{w_1u'_2}{u_1u_3}\xop_1^{-1}\zop_1 \xop_3^{-1}
  +\frac{\ka_2w_1u'_2}{u_2u_3}\zop_1\xop_2^{-1}\xop_3^{-1},
  \label{XS}\\[2mm]
\fl \lk\zop_2'\rk^{-1}&=&\frac{w'_2}{w_3}\zop_3^{-1}
-\om^{1/2}\frac{w'_2u_3}{w_1w_3}\zop_1^{-1}
   \xop_3 \zop_3^{-1}+\frac{\ka_3w'_2u_3}{w_1w_2}\zop_1^{-1}
\zop_2^{-1}\xop_3,
    \ny\\[2mm]
\fl \xop_3'&=&\frac{u_2u_3}{u_1u'_3} \xop_1^{-1}\xop_2\xop_3
-\om^{1/2} \frac{w_1u_2}{u_1u'_3}\xop_1^{-1}\zop_1 \xop_2
  +\frac{\ka_2w_1}{u'_3}\zop_1,\ny\\[2mm]
\fl \lk\zop_3'\rk^{-1}&=& \frac{\ka_1u_1w'_3}{\ka_2u_2w_3}\xop_1
\xop_2^{-1}\zop_3^{-1}
  +\frac{\ka_3w'_3}{\ka_2w_2}\zop_2^{-1}
  -\om^{1/2}\frac{\ka_1\ka_3u_1w'_3}{\ka_2u_2w_2}\xop_1
\xop_2^{-1}\zop_2^{-1}.\ny
\eea
This mapping $\xop_j^{},\zop_j^{}\mapsto\xop_j',\zop_j'$, $j=1,2,3$,
is the basic example of a class of the canonical rational mappings of the 
ultra-local Weyl algebra. The following lemma establishes the uniqueness
of the matrix structure of any such mapping.
\begin{lem}\label{lem1}
Let $\;\xop_j,\;\zop_j,\;\;j=1\ldots\Delta$ be a normalized finite-dimensional 
unitary basis (\ref{XZdef}) of the local Weyl algebra
\beq 
\xop_i\;\zop_j\;=\;q\;\zop_j\;\xop_i\;\delta_{i,j};\hs q^N=1;
       \hs\xop_j^N=\zop_j^N=1.  
\end{equation}
Let $\cal{E}$: $\hx {\xop}_j,\;{\zop}_j \mapsto
{\xop}_j',\;{\zop}_j'\;$ 
be an invertible canonical mapping in the space
of rational functions of $\xop_j,\zop_j$
such that it conserves the centers
\[ \fl {\xop\,'}_j^N\,=\,{\zop\,'}_j^N\:=\,1\,.\]
Then there exists a unique (up to a scalar multiplier)
$\;N^\Delta\times N^\Delta$ matrix $E$ such that 
for any $\Phi$ of eq.\textnormal{(\ref{r-mapping})}:
\beq \label{conjug}
\Phi(\,\xop_j',\:\zop_j')\;=\;E\:\Phi(\xop_j,\:\zop_j)\:E^{-1}. 
\end{equation}
\end{lem}
\noindent\emph{Proof}:~~ 
The ring of the rational functions of $\xop_j,\zop_j$ at root of unity
is the algebra of the polynomials of $\xop_j,\zop_j$ with 
$\mathbb{C}$-valued coefficients. Evidently this enveloping algebra
is the complete algebra of $N^\Delta\times N^\Delta$ matrices.
Since $\cal{E}$ is invertible, the envelope of $\xop_j',\zop_j'$ 
is the same matrix algebra. 
Furthermore, since $\cal{E}$ is canonical and conserves the
$N$-th powers of the Weyl elements, $\cal{E}$ is an automorphism of the
matrix algebra.
Finally, since the algebra of $N^\Delta\times N^\Delta$ matrices 
is the irreducible fundamental representation of the
semi-simple algebra $\mathfrak{gl}(N^\Delta)$, any such automorphism is 
an internal one and may be realized by the unique matrix $E$ of
(\ref{conjug}). \hfill $\square$

\subsection{Matrix part of $\;\Rop_{1,2,3}\;$ at root of unity
in terms of Fermat curve\\ cyclic functions $W_p(n)$}

Due to lemma \ref{lem1} there exists a unique (up to a scalar factor)
$N^3\times N^3$-dimensional matrix $\R_{1,2,3}$, such that
\beq\ds\R_{1,2,3}\,\xop_1^{}\;=\;\xop_1'\,\R_{1,2,3}\;,\;\;\;
\R_{1,2,3}\,\zop_1^{}\;=\;\zop_1'\,\R_{1,2,3}\;,\;\;\;
\mbox{etc.} \label{unique}
\end{equation}
for (\ref{xzprime}).

The basis independent expression for $\R_{1,2,3}$ is not a useful object,
and here we give the matrix elements of $\R_{1,2,3}$ in the basis
(\ref{XZdef}) in terms of the
Bazhanov-Baxter \cite{Bazh-Bax} cyclic functions $\:W_p(x)\;$ which we 
shall mainly use in this paper. We define
\beq\frac{W_p(n)}{W_p(n-1)}\;=\;\frac{y}{1\,-\,\om^n\,x};\hs\hs W_p(0)\:=\:1,
             \label{W-def}\end{equation}
where $\;n\in\mathbb{Z}_N\:$ and $\:p=(x,\:y)\:$ denotes a point on the
Fermat curve \beq x^N\,+\,y^N\,=\,1.\label{fermatc}\end{equation}
For $\:n>0\:$ we have
 \beq W_p(n)\;=\;\prod_{\nu=1}^n\;\frac{y}{1-\om^\nu\,x},\end{equation}
and generally $\;W_p(n+N)\;=\;W_p(n),\:$ because of
$\;\prod_{\nu=0}^{N-1}\:(1-\om^\nu\:x)\;=y^N.$
One automorphism of the Fermat curve will be important for later
calculations.
Defining $\:Op\:$ by \beq\label{O}
p=(x,y) \mapsto Op = (\om^{-1}x^{-1},\:\om^{-1/2}x^{-1}y),\end{equation}
we have
\beq\label{property} W_p(n)=\frac{1}{W_{Op}(-n)\;\Phi(n)}, \end{equation}
with
\beq\label{Phi} \Phi(n)=(-1)^n\:\om^{n^2/2}.  \end{equation}
At special points on the Fermat curve the $\:W_p(n)\:$ take simple values:
Defining \beq q_0=(0,1);\hs q_\infty=O\,q_0;\hs q_1=(\om^{-1},0);
   \label{spcq}\end{equation}
we get
\beq W_{q_0}(n)\;=\;1;\hs W_{q_\infty}(n)\;=\;\Phi^{-1}(n);\hs 1/W_{q_1}\;=
           \;\delta_{n,0}.\label{specq} \end{equation}
We now express our conjugation matrix in terms of the functions $\:W_p(n)\,$:
\begin{prop} 
In the basis (\ref{XZdef})
the matrix $\R_{1,2,3}$, solving the relations (\ref{xzprime},\ref{unique}),
 has the following matrix elements:
\bea\label{R-N-matrix} \langle i_1,i_2,i_3|\R_{1,2,3}|j_1,j_2,j_3\,\rangle\;
&\stackrel{{\rm def}}{=}& R_{i_1,i_2,i_3}^{j_1,j_2,j_3}\ny\\&=&
\delta_{i_2+i_3,j_2+j_3}\;\om^{(j_1-i_1)\,j_3}\;
{W_{p_1}(i_2-i_1)\,W_{p_2}(j_2-j_1)\over W_{p_3}(j_2-i_1)\,W_{p_4}(i_2-j_1)}\ ,
\hq\eea
where the $x$-coordinates of the four Fermat curve points are connected by
\beq \ds x_1\;x_2\;=\;\om\;\;x_3\;x_4\,.\label{consg}\end{equation}
In the terms of the variables $u_j,\; w_j,\;\ka_j,\;\;j=1,2,3$,
these points are defined by
\beq\label{x-uw}
 x_1^{}\;=\;{\om^{-1/2}\over\ka_1}\,{u_2^{}\over u_1^{}}\;,\hq
 x_2^{}\;=\;\om^{-1/2}\ka_2\,{u_2'\over u_1'}\;\hq
 x_3^{}\;=\;\om^{-1}{u_2'\over u_1^{}}\;,\hq
 x_4^{}\;=\;\om^{-1}{\ka_2\over\ka_1}\,{u_2^{}\over u_1'}\;,
\end{equation}
\beq\label{y-uw}
{y_3\over y_1}\;=\;\ka_1\,{w_1^{}\over u_3'}\;,\hq
{y_4\over y_1}\;=\;\om^{-1/2}\ka_3\,{w_3^{}\over w_2}\;,\hq
{y_3\over y_2}\;=\;{w_2'\over w_3}\;,\hq
{y_4\over y_2}\;=\;\om^{-1/2}{\ka_3\over\ka_1}
\,{u_3'\over w_1'}\;,\end{equation}
where the $\;u_j',\;w_j'$ and $\;u_j,\;w_j$ are related by the functional
transformation (\ref{fctm}):
\beq u_j'=\Rop^{(f)}_{1,2,3}\circ u_j^{},\hs w_j'=\Rop^{(f)}_{1,2,3}\circ w_j.
          \label{us-f}\end{equation}
\end{prop}
\noindent
{\it Proof.}
We shall give the proof, that (\ref{R-N-matrix}) produces the rational 
mapping, 
for the first line of (\ref{XS}) only, the other equations follow analogously. 
\\ First observe that the matrix elements of the operator $\R_{1,2,3}$ satisfy
several recurrent relations. In particular, we will need the recursion
\[ \fl R^{j_1,j_2,j_3}_{i_1,i_2+1,i_3-1}=
R^{j_1,j_2,j_3}_{i_1,i_2,i_3}\cdot \frac{y_1}{y_4}\cdot
\frac{1-\om^{i_2-j_1+1}x_4}{1-\om^{i_2-i_1+1}x_1}\]
which can be rewritten in the form
\beq    R^{j_1,j_2,j_3}_{i_1,i_2,i_3}\;\om^{-j_1}\,=\,
\frac{1}{\om^{i_2+1}x_4}R^{j_1,j_2,j_3}_{i_1,i_2,i_3}
 +\frac{x_1 y_4}{\om^{i_1}y_1 x_4}R^{j_1,j_2,j_3}_{i_1,i_2+1,i_3-1}
  -\frac{y_4}{\om^{i_2+1}x_4 y_1}R^{j_1,j_2,j_3}_{i_1,i_2+1,i_3-1}\ .
\end{equation}
However, this recurrent relation is the matrix element
$\langle i_1i_2i_3|\cdot|j_1j_2j_3 \rangle$
of the operator equality
\[ \fl \R_{1,2,3}\cdot \lk\xop_1'\rk^{-1}=\lk\frac{1}{\om\, x_4}\xop_2^{-1}
  +\frac{x_1 y_4}{y_1 x_4}\xop_1^{-1}\zop_2^{-1}\zop_3
  -\frac{y_4}{\om\:x_4 y_1}\xop_2^{-1}\zop_2^{-1}\zop_3\rk\cdot\R_{1,2,3}.\]
This coincides with the first line in (\ref{XS}), provided
the identification (\ref{x-uw}) and (\ref{y-uw}) is valid for the 
Fermat points $\:(x_1,y_1)$ and $\;(x_4,y_4)$.\hfill $\square$
\begin{remark}\label{rem-par}
$\hx\R$ is a matrix function of three continuous parameters 
$\;x_1,\,x_2,\,x_3$ 
and three discrete parameters: the phases of $\;y_1,\,y_2,\,y_3$.
Equivalently, one may use $\ds\kappa_1\frac{u_1}{u_2}$,
$\ds \kappa_3\frac{w_3}{w_2}$ and $\ds\frac{w_1}{u_3}$ as the 
continuous parameters and the phases of $\:u_1',\, u_2',\, w_1'$
as the discrete parameters. Formulas (\ref{x-uw},\ref{y-uw}) 
establish the correspondence between these choices. We call the 
parameterization of $\R_{1,2,3}$ in the terms of $\:u_j,w_j,\kappa_j\:$ 
a "free parameterization".
\end{remark}
\begin{remark}\label{rrfr}
$\hx$Formulated in terms of mappings, the automorphism $\Rop_{1,2,3}$ of the 
ultra-local Weyl algebra at the root of unity is presented as the 
superposition of a pure functional mapping and the finite dimensional 
similarity transformation:
\beq  \ds \Rop_{1,2,3}^{}\circ\Phi\;=\;
\R_{1,2,3}^{}\;\lk\Rop^{(f)}_{1,2,3}\circ\Phi\rk\;\R_{1,2,3}^{-1}\:.
\label{frfr}\end{equation}
\end{remark}

\subsection{Free bosonic realization of $\R_{1,2,3}$}

Analogous to the continuum realization of the TE-Boltzmann weights proposed 
in eq.(4.1) of \cite{Bazh-Bax-st-tri}, also our (\ref{R-N-matrix}) can be
written in a bosonic realization. In this realization the cyclic weights
$\;W_p(n)$ of (\ref{W-def}) are replaced by the following Gaussian weights:
\beq 
W_x(\si)\;=\;\exp{\lk\frac{i}{2\hbar}\:\frac{x}{x-1}\;\si^2\rk}\,;\hx
\hs\sig\in\Rer,\hq x\in{\mathbb{C}};\hq {\Im}m\,\frac{x}{x-1}>0.
\label{Gauss}\end{equation}    
At each vertex $j$ of a graph define a pair of operators 
$\:\qop_j,\;\pop_j\:$ satisfying 
$\:[\,\qop_{j'},\pop_j\:]=i\hbar\delta_{j,j'}$, 
and scalar variables $u_j,\;w_j$. We choose a 
basis $\;|\,\sig_j\rangle\;$ with $\langle \sig_{j'}|\,\sig_j\rangle=
   \delta(\sig_{j'}-\sig_j),$ such that for $\:\psi(\sig)\in L^2\:$ we have
\beq  
\psi(\sig_j)\stackrel{def}{=}
\langle \sig_j|\psi\rangle;\hs
  \langle\sig_j|\:\qop_j\,|\psi\rangle=\sig_j\psi(\sig_j);\hs
  \langle\sig_j|\:\pop_j\,|\psi\rangle=\frac{\hbar}{i}\:
  \frac{\partial\psi(\sig_j)}{\partial\sig_j}. \label{bobas}\end{equation}
For each set of vertices we define the corresponding operators and scalars and
a direct product space of the single vertex spaces.
Consider now the following mapping $\R_{1,2,3}$ in the product space of three 
points $\:j=1,2,3$ by  
\bea \R_{123}\:\qop_j\:\R_{123}^{-1}&=&\sum_{k=1}^{3}
\frac{\partial \log{u_j'}}
{\partial \log{u_k}}\:\qop_k\;+\;\frac{\partial\log{u_j'}}
{\partial \log{w_k}}\:\pop_k\;\equiv\;\qop_j'\ny\\
\fl \R_{123}\:\pop_j\:\R_{123}^{-1}&=&\sum_{k=1}^{3}
\frac{\partial\log{w_j'}}
{\partial\log{u_k}}\:\qop_k\;+\;\frac{\partial\log{w_j'}}
{\partial\log{w_k}}\:\pop_k\;\equiv\;\pop_j'.\label{trsi}\eea
where the relation between the primed and unprimed scalars $u_j,\;w_j$
is given by (\ref{fctm}), putting there formally $\:N=1\,.$ The $\ka_j$ 
are further parameters
(``coupling constants``) at the vertices $j$. 
As the following proposition shows, equations (\ref{trsi}) are the 
bosonic continuum analogs to eqs.(\ref{XS}) of the discrete case.
\begin{prop} In the basis (\ref{bobas}) the operator $\R_{1,2,3}$ has the 
following kernel
\beq\label{bR-N-matrix}
\ds\begin{array}{clc} &\ds
\langle\sigma_1^{},\sigma_2^{},\sigma_3^{}| \R_{1,2,3}
|\si'_1,\si'_2,\si'_3\rangle &\\[3mm]
&\ds \hs =\; \delta(\si_2^{}+\si_3^{}-\si'_2-\si'_3)\;
e^{-\frac{i}{\hbar}(\sigma'_1-\sigma^{}_1)\,\sigma'_3}\;
\frac{W_{x_1}(\sigma^{}_2-\sigma^{}_1)\,W_{x_2}(\sigma'_2-\sigma'_1)}
{W_{x_3}(\sigma'_2-\sigma^{}_1)\,W_{x_4}(\sigma^{}_2-\sigma'_1)}\;,
\end{array}\label{bobo}\end{equation}
with the constraint $\; x_1\:x_2\:=\:x_3\:x_4\,.$
In terms of the variables $u_j,\:w_j,\:\ka_j,\:\,(j=1,2,3)$, the 
$x_k\;\in\;\Rer\;$ are defined by $\:$ 
(obtained by putting formally $\om^{-1/2}\rightarrow-1$ in (\ref{x-uw})):
\beq\label{bx-uw}
x_1^{}\;=\,-\frac{1}{\ka_1}\,\frac{u_2^{}}{u_1^{}}\;;\hq
x_2^{}\;=\,-\ka_2\,\frac{u_2'}{u_1'}\;;\hq
x_3^{}\;=\:\frac{u_2'}{u_1^{}}\;;\hq
x_4^{}\;=\:\frac{\ka_2}{\ka_1}\,\frac{u_2^{}}{u_1'}\,,\end{equation}
where $\;u_1'$ and $u_2'$ are defined as in (\ref{us-f}) with $N=1$.
\end{prop}
\noindent\emph{Proof}:~~ 
We give the proof for one of the six equations (\ref{trsi}), as the other
equations follow analogously. Let us write shorthand $\;|\sig\rangle\:$ for 
$\;|\si_1,\si_2,\si_3\rangle\:$ and $\:d^3\sig\:=\:d\si_1d\si_2d\si_3\:$
etc. We consider:
\beq \int d^3\si'\;\langle\:\si\:|\:\R_{1,2,3}\,|\:\si'\:\rangle 
\langle\: \si'\,|\:\qop_3\,|\:\si''\,\rangle\:=\:
\int d^3\si'\;\langle\,\si\,|\,\qop_3'\,|
\,\si'\,\rangle \langle \,\si'\,|\,\R_{1,2,3}\,|\,\si''\rangle
\label{exqd}\end{equation} 
which should be satisfied for all 
$\;\si_1,\si_2,\si_3,\si_1'',\si_2'',\si_3''$.
Written more explicitly, the kernel (\ref{bR-N-matrix}) is:
\[ \fl R_{\si_1,\si_2,\si_3}^{\si'_1,\si'_2,\si'_3}\;=\;
\delta(\si_2+\si_3-\si'_2-\si'_3)\;
\exp{\lk\frac{i}{2\,\hbar}\:\Sigma(\si,\si')\rk}; \]
where
\bea \Sigma(\si,\si')&=&
\frac{w_1(\si_2-\si_1)^2\,+u_3(\si_1-\si_2')^2}{u_2^{-1}w_1(\ka_1u_1+u_2)}+
   \frac{(\ka_1u_1w_2+\ka_3u_2w_3+\ka_1\ka_3u_1w_3)(\si_2-\si_1')^2}
   {\ka_3w_3(\ka_1u_1+u_2)}\ny\\\fl &&\hspace{-2mm}+\;
 \frac{u_3(\ka_1u_1w_2+\ka_3u_2w_3+\ka_1\ka_3u_1w_3)(\si_2'-\si_1')^2}
 {(\ka_3u_3w_3+w_1w_2+u_3w_2)(\ka_1u_1+u_2)}\; -\,2(\si_1'-\si_1)\si_3'.\eea
From (\ref{trsi}) we get: 
\[\fl \qop_3'=\frac{(w_1+u_3)u_2(\qop_2-\qop_1)+u_2u_3\qop_3
          +(u_2+\ka_1u_1)w_1\pop_1}{u_2u_3+u_2w_1+\ka_1u_1w_1}. \]
and (\ref{exqd}) becomes:
\bea \lefteqn{\int d^3\sig'\;\delta(\si_2+\si_3-\si'_2-\si'_3)
\exp{\lk\frac{i}{2\hbar}\Sigma(\sig,\sig')\rk}\;\sig_3''\;
   \delta^3(\si'-\si'')}\ny\\
\fl &=&\int d^3\sig'\;\delta^3(\si-\si')\:
\frac{(w_1+u_3)u_2(\si_2'-\si_1')+u_2u_3\si_3'
  +(u_2+\ka_1u_1)w_1\frac{\hbar}{i}\frac{\partial}{\partial \si_1'}}
 {u_2u_3+u_2w_1+\ka_1u_1w_1}\;\times\ny\\ \fl &&\hspace{1cm}
 \times\:\delta(\si_2'+\si_3'-\si_2''-\si_3'')
 \exp{\lk\frac{i}{2\hbar}\Sigma(\sig',\sig'')\rk}.\label{peq}\eea     
Since 
\[\fl (u_2+\ka_1u_1)\frac{w_1}{2}\,\frac{\partial\Sigma(\sig',\si'')}
 {\partial\si_1'}\;=\;(w_1+u_3)u_2\sig_1'-u_2w_1\sig_2'-u_2u_3\sig_2''
    +(w_1u_2+\ka_1u_1w_1)\sig_3'',\]  eq.(\ref{peq}) reduces to
\[\fl \si_3''\delta(\si_2+\si_3-\si_2''-\si_3'')=
\frac{u_2u_3(\sig_2+\si_3-\si_2'')+(u_2w_1+\ka_1u_1w_1)\si_3''}
{u_2u_3+u_2w_1+\ka_1u_1w_1}\delta(\si_2+\si_3-\si_2''-\si_3'')\]
\hfill $\square$

\section{The modified tetrahedron equation}\label{moteeq}

\begin{figure}[t]
\setlength{\unitlength}{0.014mm}
\begin{center}
{\renewcommand{\dashlinestretch}{30}
\begin{picture}(9037,7959)(0,-10)
\path(691,6267)(1366,7842)
\path(1346.304,7719.885)(1366.000,7842.000)(1291.155,7743.520)
\path(1591,6267)(916,7842)
\path(990.845,7743.520)(916.000,7842.000)(935.696,7719.885)
\path(2491,6942)(16,7167)
\path(138.223,7186.013)(16.000,7167.000)(132.791,7126.259)
\path(253,6247)(2503,7597)
\path(2415.536,7509.536)(2503.000,7597.000)(2384.666,7560.985)
\put(810,6910){\makebox(0,0)[lb] {$\skk2$}}
\put(1070,7500){\makebox(0,0)[lb]{$\skk3$}}
\put(1250,7122){\makebox(0,0)[lb]{$\skk1$}}
\put(1850,7038){\makebox(0,0)[lb]{$\skk4$}}
\put(1400,6770){\makebox(0,0)[lb]{$\skk5$}}
\put(900,6447){\makebox(0,0)[lb] {$\skk6$}}
\path(1636,5952)(1636,5277)
\path(1606.000,5397.000)(1636.000,5277.000)(1666.000,5397.000)
\put(1800,5450){\makebox(0,0)[lb]{$\sk{\Rop_{1,4,5}}$}}
\path(6856,5907)(6856,5232)
\path(6826.000,5352.000)(6856.000,5232.000)(6886.000,5352.000)
\put(5900,5450){\makebox(0,0)[lb]{$\sk{\Rop_{2,4,6}}$}}
\path(6541,3072)(5641,4872)
\path(5721.498,4778.085)(5641.000,4872.000)(5667.833,4751.252)
\path(5568,3433)(7818,4783)
\path(7730.536,4695.536)(7818.000,4783.000)(7699.666,4746.985)
\path(7477,3414)(5227,4539)
\path(5347.748,4512.167)(5227.000,4539.000)(5320.915,4458.502)
\path(6244,3063)(7144,4863)
\path(7117.167,4742.252)(7144.000,4863.000)(7063.502,4769.085)
\put(6325,3050){\makebox(0,0)[lb]{$\skk3$}}
\put(5920,3792){\makebox(0,0)[lb]{$\skk5$}}
\put(5970,4242){\makebox(0,0)[lb]{$\skk1$}}
\put(6630,3610){\makebox(0,0)[lb]{$\skk2$}}
\put(6840,4000){\makebox(0,0)[lb]{$\skk6$}}
\put(6320,4040){\makebox(0,0)[lb]{$\skk4$}}
\path(1726,3207)(2401,4782)
\path(2381.304,4659.885)(2401.000,4782.000)(2326.155,4683.520)
\path(2839,3176)(2164,4751)
\path(2238.845,4652.520)(2164.000,4751.000)(2183.696,4628.885)
\path(1260,3233)(3510,4583)
\path(3422.536,4495.536)(3510.000,4583.000)(3391.666,4546.985)
\path(3526,3653)(1051,3878)
\path(1173.223,3897.013)(1051.000,3878.000)(1167.791,3837.259)
\put(2000,4422){\makebox(0,0)[lb]{$\skk3$}}
\put(1800,3850){\makebox(0,0)[lb]{$\skk2$}}
\put(2470,4062){\makebox(0,0)[lb]{$\skk5$}}
\put(2700,3560){\makebox(0,0)[lb]{$\skk1$}}
\put(2130,3580){\makebox(0,0)[lb]{$\skk4$}}
\put(1600,3250){\makebox(0,0)[lb]{$\skk6$}}
\path(1636,2892)(1636,2217)
\path(1606.000,2337.000)(1636.000,2217.000)(1666.000,2337.000)
\put(1800,2430){\makebox(0,0)[lb]{$\sk{\Rop_{2,4,6}}$}}
\path(225,353)(2475,1703)
\path(2387.536,1615.536)(2475.000,1703.000)(2356.666,1666.985)
\path(876,235)(1551,1810)
\path(1531.304,1687.885)(1551.000,1810.000)(1476.155,1711.520)
\path(2487,368)(12,593)
\path(134.223,612.013)(12.000,593.000)(128.791,552.259)
\path(1780,215)(1105,1790)
\path(1179.845,1691.520)(1105.000,1790.000)(1124.696,1667.885)
\put(1000,957){\makebox(0,0)[lb]{$\skk6$}}
\put(400,600){\makebox(0,0)[lb]{$\skk4$}}
\put(980,250){\makebox(0,0)[lb]{$\skk2$}}
\put(1500,912){\makebox(0,0)[lb]{$\skk5$}}
\put(1250,1450){\makebox(0,0)[lb]{$\skk3$}}
\put(1680,470){\makebox(0,0)[lb]{$\skk1$}}
\path(6856,2892)(6856,2217)
\path(6826.000,2337.000)(6856.000,2217.000)(6886.000,2337.000)
\put(5900,2430){\makebox(0,0)[lb]{$\sk{\Rop_{1,4,5}}$}}
\path(4696,6132)(3796,7932)
\path(3876.498,7838.085)(3796.000,7932.000)(3822.833,7811.252)
\path(3796,6132)(4696,7932)
\path(4669.167,7811.252)(4696.000,7932.000)(4615.502,7838.085)
\path(5821,6582)(3571,7707)
\path(3691.748,7680.167)(3571.000,7707.000)(3664.915,7626.502)
\path(3571,6357)(5821,7707)
\path(5733.536,7619.536)(5821.000,7707.000)(5702.666,7670.985)
\put(4470,6690){\makebox(0,0)[lb]{$\skk5$}}
\put(4950,7050){\makebox(0,0)[lb]{$\skk4$}}
\put(4470,7280){\makebox(0,0)[lb]{$\skk2$}}
\put(4060,6520){\makebox(0,0)[lb]{$\skk6$}}
\put(4020,7500){\makebox(0,0)[lb]{$\skk1$}}
\put(4070,6987){\makebox(0,0)[lb]{$\skk3$}}
\path(8089,66)(7189,1866)
\path(7269.498,1772.085)(7189.000,1866.000)(7215.833,1745.252)
\path(7711,102)(8611,1902)
\path(8584.167,1781.252)(8611.000,1902.000)(8530.502,1808.085)
\path(6628,496)(8878,1846)
\path(8790.536,1758.536)(8878.000,1846.000)(8759.666,1809.985)
\path(9025,165)(6775,1290)
\path(6895.748,1263.167)(6775.000,1290.000)(6868.915,1209.502)
\put(8500,1450){\makebox(0,0)[lb]{$\skk6$}}
\put(8100,650){\makebox(0,0)[lb]{$\skk2$}}
\put(7820,120){\makebox(0,0)[lb]{$\skk3$}}
\put(7100,910){\makebox(0,0)[lb]{$\skk4$}}
\put(7580,1180){\makebox(0,0)[lb]{$\skk5$}}
\put(7590,690){\makebox(0,0)[lb]{$\skk1$}}
\path(4606,12)(5506,1812)
\path(5479.167,1691.252)(5506.000,1812.000)(5425.502,1718.085)
\path(5506,12)(4606,1812)
\path(4686.498,1718.085)(4606.000,1812.000)(4632.833,1691.252)
\path(3481,462)(5731,1587)
\path(5637.085,1506.502)(5731.000,1587.000)(5610.252,1560.167)
\path(5731,237)(3481,1587)
\path(3599.334,1550.985)(3481.000,1587.000)(3568.464,1499.536)
\put(4170,920){\makebox(0,0)[lb]{$\skk4$}}
\put(5140,867){\makebox(0,0)[lb]{$\skk3$}}
\put(5100,1400){\makebox(0,0)[lb]{$\skk6$}}
\put(4650,1150){\makebox(0,0)[lb]{$\skk5$}}
\put(5120,380){\makebox(0,0)[lb]{$\skk1$}}
\put(4700,640){\makebox(0,0)[lb]{$\skk2$}}
\path(8116,6132)(7216,7932)
\path(7296.498,7838.085)(7216.000,7932.000)(7242.833,7811.252)
\path(7585,6060)(8485,7860)
\path(8458.167,7739.252)(8485.000,7860.000)(8404.502,7766.085)
\path(8881,6762)(6631,7887)
\path(6751.748,7860.167)(6631.000,7887.000)(6724.915,7806.502)
\path(3481,7032)(2806,7032)
\path(2926.000,7062.000)(2806.000,7032.000)(2926.000,7002.000)
\path(6181,7032)(6856,7032)
\path(6736.000,7002.000)(6856.000,7032.000)(6736.000,7062.000)
\path(2671,957)(3346,957)
\path(3226.000,927.000)(3346.000,957.000)(3226.000,987.000)
\path(6496,957)(5821,957)
\path(5941.000,987.000)(5821.000,957.000)(5941.000,927.000)
\path(6611,6075)(8861,7425)
\path(8773.536,7337.536)(8861.000,7425.000)(8742.666,7388.985)
\put(8000,7240){\makebox(0,0)[lb]{$\skk2$}}
\put(7450,7480){\makebox(0,0)[lb]{$\skk1$}}
\put(8430,7000){\makebox(0,0)[lb]{$\skk4$}}
\put(7550,6750){\makebox(0,0)[lb]{$\skk5$}}
\put(7800,6300){\makebox(0,0)[lb]{$\skk3$}}
\put(8000,6700){\makebox(0,0)[lb]{$\skk6$}}
\put(2700,7150){\makebox(0,0)[lb]{$\sk{\Rop_{1,2,3}}$}}
\put(6000,7150){\makebox(0,0)[lb]{$\sk{\Rop_{3,5,6}}$}}
\put(2600,1070){\makebox(0,0)[lb]{$\sk{\Rop_{3,5,6}}$}}
\put(5800,1070){\makebox(0,0)[lb]{$\sk{\Rop_{1,2,3}}$}}
\put(0,1270){\makebox(0,0)[lb]{$Q_4$}}
\put(3950,1750){\makebox(0,0)[lb]{$Q_5$}}
\put(8950,900){\makebox(0,0)[lb]{$Q_6$}}
\put(300,3900){\makebox(0,0)[lb]{$Q_3$}}
\put(7750,3900){\makebox(0,0)[lb]{$Q_7$}}
\put(0,7800){\makebox(0,0)[lb]{$Q_2$}}
\put(5000,7800){\makebox(0,0)[lb]{$Q_1$}}
\put(8950,7800){\makebox(0,0)[lb]{$Q_8$}}
\end{picture}
}
\caption{\footnotesize{Graphical image of the two equivalent ways of 
transforming the four-line-graph ("quadrilateral") $Q_1$ into graph 
$Q_5$, which leads to tetrahedron equation. Observe that each graph 
contains only two triangles which can be transformed by a mapping 
$\Rop$: In graph $Q_1$ either the line 124 can be moved 
downward through the point 3 (leading to graph $Q_2$), or the line 456 
can be moved upward through point 3 (leading to $Q_8$). Both the left 
hand and right hand sequences of four transformations lead to the same 
graph $Q_5.$}} \label{tet-graph}
\end{center}\end{figure}
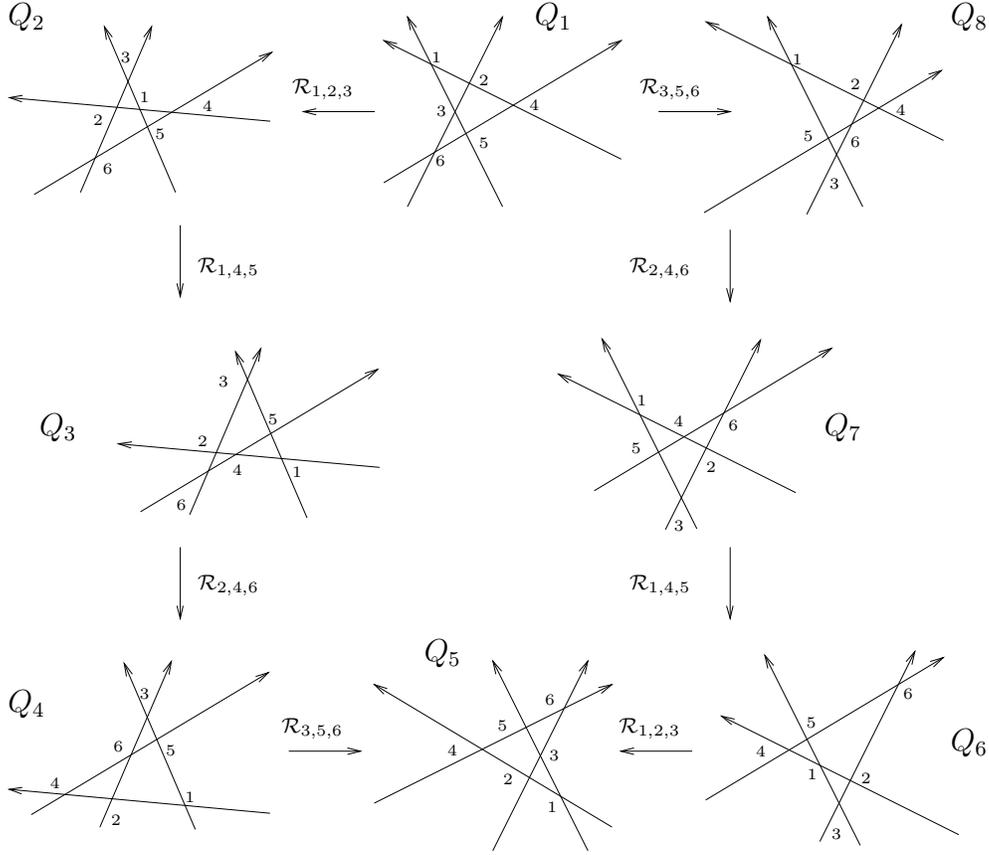

For three-dimensional integrable spin models the Tetrahedron Equation (TE) 
plays a role which the Yang-Baxter
equation has for two-dimensional integrable spin models. The TE provides the 
commutativity of so-called layer-to-layer transfer matrices. In our case, 
where the dynamical variables form an affine Weyl algebra, 
we are able to define a more general equation: the Modified Tetrahedron 
Equation (MTE), which provides the commutativity of more complicated transfer
matrices, see e.g \cite{bms-mte,ms-mte}.
In Fig.~\ref{tet-graph} we show a graphical image of the mappings leading to 
the tetrahedron equation (\ref{tet-prop}): the sequence of mappings  
$Q_1\rightarrow 
Q_2\rightarrow Q_3\rightarrow Q_4\rightarrow Q_5$ gives the same $Q_5$ 
as $Q_1\rightarrow Q_8\rightarrow Q_7\rightarrow Q_6\rightarrow Q_5$.

Following eq.(\ref{rma}) let us fix the following
notation for the superposition of two mappings $\mathcal{A}$ and
$\mathcal{B}$:
\beq \ds\lk\lk\mathcal{A}\cdot\mathcal{B}\rk\circ\Phi\rk\;
\stackrel{def}{=}\;
\lk\mathcal{A}\circ\lk\mathcal{B}\circ\Phi\rk\rk\;.
\end{equation}
Due to the uniqueness of the mapping $\Rop$ discussed in 
section \ref{R-mapping},
p.\pageref{R-mapping}, we arrive at the 
\begin{prop}\label{tet-prop}
The mapping $\Rop$ solves the Tetrahedron Equation:  
\beq\label{op-te}
\ds \Rop_{123}\cdot\Rop_{145}\cdot\Rop_{246}\cdot\Rop_{356}\;=\;
\Rop_{356}\cdot\Rop_{246}\cdot\Rop_{145}\cdot\Rop_{123}\,,\label{tetra}
\end{equation} acting in the space of the twelve {\it affine} Weyl
elements $\;\uop_1,\wop_1,\uop_2,\wop_2,\ldots,\wop_5,\uop_6,\wop_6.$
\end{prop}
Since, as has been discussed in Sec. \ref{fuN}, any rational automorphism
of the ultra-local Weyl algebra implies a rational mapping in the space
of $N$-$th$ powers of the parameters of the representation, its is a direct 
consequence of (\ref{op-te}) that the $\Rop^{(f)}_{i,j,k}$ of (\ref{fctm}) 
solve the tetrahedron equation with the variables 
$\:u_j^N,\;w_j^N,\;j=1,\ldots,8$: \\[-4mm]
\beq\label{fc-te}\ds \Rop^{(f)}_{123}\,\cdot\,\Rop^{(f)}_{145}\,\cdot\,
\Rop^{(f)}_{246}\,\cdot\,\Rop^{(f)}_{356}\;=\;\Rop^{(f)}_{356}\,\cdot\,
\Rop^{(f)}_{246}\,\cdot\,\Rop^{(f)}_{145}\,\cdot\,\Rop^{(f)}_{123}\,.
\end{equation} 
We want to get this {\it functional} tetrahedron equation not only for
the $N$-$th$ powers of the variables but for the variables $u_j$ and $w_j$ 
directly. However, when taking the $N-th$ roots, not all phases of the 
$u_j,\;w_j$ can be chosen independently. In Sec. \ref{disph} we shall show 
explicitly how to make an independent choice of phases.

Once an appropriate choice of phases has  been made, we obtain the 
{\it functional} tetrahedron equation on the variables $u_j,\;w_j$. 
Now using (\ref{frfr}) the functional tetrahedron equation can be 
canceled between the two sides of (\ref{op-te}), and we are left with
the {\it Modified Tetrahedron Equation} 
for the {\it finite dimensional} $\R$-matrices:
\beq\label{mte-general}
\ds\begin{array}{l}\ds\R_{1,2,3}
\cdot\lk\Ropf_{1,2,3}\circ\R_{1,4,5}\rk
\!\cdot\!\lk\Ropf_{1,2,3}\Ropf_{1,4,5}\circ\R_{2,4,6}\rk
\!\cdot\!\lk\Ropf_{1,2,3}\Ropf_{1,4,5}\Ropf_{2,4,6}\circ\R_{3,5,6}\rk
\\[4mm]
\sim\; \R_{3,5,6}\cdot\lk\Ropf_{3,5,6}\circ\R_{2,4,6}\rk
\!\cdot\!\lk\Ropf_{3,5,6}\Ropf_{2,4,6}\circ\R_{1,4,5}\rk
\!\cdot\!\lk\Ropf_{3,5,6}\Ropf_{2,4,6}\Ropf_{1,4,5}\circ\R_{1,2,3}\rk
\end{array}\end{equation}
Observe that due to the cancellation of the functional tetrahedron equation
there is no $\Ropf_{3,5,6}$  on the left hand side of (\ref{mte-general}) 
and no $\Ropf_{1,2,3}$ on the right hand side. 
Because of the uniqueness of the mapping shown in Lemma \ref{lem1}
on page \pageref{lem1}, the left and right hand sides of (\ref{mte-general}) 
may differ only by a scalar factor, which arises
when we pass from the equivalence of the mappings to the equality
of the matrices.
So, in matrix element notation the MTE reads
\beq\label{mte-elements}\ds\begin{array}{l}\ds \sum_{j_1...j_6}\;
\lk R^{(1)}\rk_{i_1,i_2,i_3}^{j_1,j_2,j_3}\,
\lk R^{(2)}\rk_{j_1,i_4,i_5}^{k_1,j_4,j_5}\,
\lk R^{(3)}\rk_{j_2,j_4,i_6}^{k_2,k_4,j_6}\,
\lk R^{(4)}\rk_{j_3,j_5,j_6}^{k_3,k_5,k_6}\\[4mm]
\ds\hs\hq\;=\;\rho\,\sum_{j_1...j_6}\;
\lk R^{(8)}\rk_{i_3,i_5,i_6}^{j_3,j_5,j_6}\,
\lk R^{(7)}\rk_{i_2,i_4,j_6}^{j_2,j_4,k_6}\,
\lk R^{(6)}\rk_{i_1,j_4,j_5}^{j_1,k_4,k_5}\,
\lk R^{(5)}\rk_{j_1,j_2,j_3}^{k_1,k_2,k_3}\;,
\end{array}\end{equation}
where $R^{(1)}$ corresponds to $\R_{1,2,3}$, $R^{(2)}$ to
$\Ropf_{1,2,3}\circ\R_{1,4,5}$, etc.
Here $\rho$ is the scalar factor. The matrix elements of each
$R^{(j)}$ as functions of the Fermat-curve parameters $x_i^{(j)},\,y_i^{(j)}$
are given by Proposition 3 with functional mappings
applied as shown in (\ref{mte-general}).
The $N^3$-$th$ power of the scalar factor in (\ref{mte-elements}) is
\beq  \rho^{N^3}\;=\;
\frac{\ds \det\R^{(1)}\ \det\R^{(2)}\ \det\R^{(3)}\ \det\R^{(4)}}
{\ds \det \R^{(8)}\ \det \R^{(7)}\ \det \R^{(6)}\ \det \R^{(5)}},
\label{rrho}\end{equation}
and this can be obtained from the determinant of one single matrix
$\;\R_{1,2,3}\;$ just by substituting the respective coordinates.

If the realization of the mapping by the free bosonic weight function 
(\ref{bobo}) is used, 
the corresponding modified tetrahedron equation involves integrations over 
$\;\mathbb{R}\;$ instead of the summations over $\;\mathbb{Z}_N$. The bosonic 
MTE may be proven directly with the help of Gaussian integrations.

\subsection{Calculation of the determinant of $\;\R_{1,2,3}$}

We use the representation (\ref{R-N-matrix}) to find a closed
expression for $\;\det\:\R_{1,2,3}$. The numerator term
$\;W_{p_2}(j_2-j_1)\;$ is diagonal, so it just
contributes a factor $\;\lk\prod_n W_{p_2}(n)\rk^{N^2}$ to the determinant.
For later convenience, we treat the other numerator factor
$\;W_{p_1}(i_2-i_1)\;$  of (\ref{R-N-matrix}) differently: using the Fermat 
curve automorphism (\ref{O}) we write
\[ \fl W_{p_1}(n)=\frac{1}{W_{Op_1}(-n)\;\Phi(n)}\,. \]
$W_{Op_1}(i_1-i_2)$ is diagonal and its determinant is trivially
calculated. We combine the factor $\Phi(i_2-i_1)$ with the two non-diagonal
terms $W_{p_3},\;W_{p_4}$ and write:
\beq \det\langle i_1,i_2,i_3|\,\R|j_1,j_2,j_3 \rangle
 =\lk\prod_{n=0}^{N-1}\frac{W_{p_2}(n)}{W_{Op_1}(n)}\rk^{\!\!N^2}\!\!\det
\frac{\delta_{i_2+i_3,j_2+j_3}\;\om^{(j_1-i_1)\,j_3}}
{\Phi(i_2-\!i_1)\,W_{p_3}(j_2-\!i_1)\,W_{p_4}(i_2-\!j_1)}
               \label{njk}\end{equation}
We now calculate the determinant on the right hand side of (\ref{njk}) 
from its finite Fourier transform in the indices $i_1$ and $j_1$. 
So we define and evaluate
\bea \lefteqn{\langle i_1,i_2,i_3|\,\R'_{1,2,3}|j_1,j_2,j_3\rangle}\ny\\
\fl  &=&\delta_{i_2+i_3,j_2+j_3}\,\frac{1}{N}\,
     \sum_{a,b\in \mathbb{Z}_N}\:\om^{i_1a-j_1b}
\frac{\om^{(b-a)\,j_3}}{\Phi(i_2-a)\,W_{p_3}(j_2-a)\,W_{p_4}(i_2-b)}
\ny\\ 
\fl &=&\delta_{i_2+i_3,j_2+j_3}\,\frac{1}{N}\,
\sum_{a',b'}\:\om^{i_1(j_2-a')-j_1(i_2-b')-(i_2-j_2)a'}
\frac{\om^{(i_2-j_2+a'-b')\,j_3}}{\Phi(i_2-j_2)\,\Phi(a')\,
           W_{p_3}(a')\,W_{p_4}(b')}\ny\\
\fl &=&\delta_{i_2+i_3,j_2+j_3}\,
  \frac{\om^{i_1j_2-i_2j_1+(j_3-i_3)j_3}}{\Phi(j_3-i_3)}\;
\frac{1}{N}\,\sum_{a'}\:\frac{\om^{(i_3-i_1)a'}}{\Phi(a')\,W_{p_3}(a')}
\,\sum_{b'}\:\frac{\om^{-(j_3-j_1)b'}}{W_{p_4}(b')}\label{four}\eea
In the last line we have used the property
\beq \Phi(a+b)=\Phi(a)\:\Phi(b)\:\om^{ab}\label{fak}\end{equation}
which, supplying a factor $\;\om^{(i_2-j_2)a'},\:$ decouples the
$a'$- and $b'$-summations in the presence of $\;\delta_{i_2+i_3,j_2+j_3}$.
The determinant of the first three factors of the last line of (\ref{four})
will again be calculated from its Fourier transform. So we define and evaluate
\bea \langle i_1,i_2,i_3|\,P\,|j_1,j_2,j_3 \rangle&=&
 \frac{1}{N^2}\sum_{c,d}\,\om^{i_1c-j_1d}\delta_{i_2+i_3,j_2+j_3}\,
 \frac{\om^{j_2c-i_2d+(j_3-i_3)j_3}}{\Phi(i_3)\Phi(j_3)}\;\om^{i_3j_3}\ny\\
&=& \delta_{i_2+i_3,j_2+j_3}\,\delta_{i_1,-j_2}\,\delta_{j_1,-i_2}\,
 \frac{\Phi(j_3)}{\Phi(i_3)},\label{comc}\eea
so that \beq \det \,\langle i_1,i_2,i_3|\,P\,|j_1,j_2,j_3 \rangle\;=
\;\det\;||\:\delta_{i_2+i_3,j_2+j_3}\,\delta_{i_1,-j_2}\,\delta_{j_1,-i_2}\,||
   \;=\;(-1)^{N^2(N-1)/2}.\label{comd}\end{equation}
Combining (\ref{njk}),(\ref{four}),(\ref{comc}) and (\ref{comd}), 
our preliminary result is:
\bea \lefteqn{\det <i_1,i_2,i_3|\R_{1,2,3}|j_1,j_2,j_3>}\ny\\
\fl &=&\left((-1)^{(N-1)/2}\:\prod_k\frac{W_{p_2}(k)}{W_{Op_1}(k)}
\;\prod_m\sum_{a}\frac{\om^{ma}}{\Phi(a)W_{p_3}(a)}
\;\prod_n\sum_{b}\frac{\om^{nb}}{W_{p_4}(b)}\right)^{\!\!N^2}.
           \label{det-R-prel}\eea
We may simplify this formula by introducing the function $V(x)$ on the 
Fermat curve
\beq V(x)\;\stackrel{{\rm def}}{=}\;
\prod_{n=1}^{N-1}\;(1\:-\,\omega^{n+1}\,x)^n.\label{dgam}\end{equation}
Writing $\;\lk V(\om^{-1})\rk^2$ as product of $N^2$ factors of the
form $\:(1\,-\om^n)\:$ times powers of $\,\om$, we compute
\beq V(\om^{-1})\;=
              \;N^{N/2}\:e^{i\pi(N-1)(N-2)/12}.\label{Vom}\end{equation}
It is useful to note that
\beq \prod_n\sum_a\,\frac{\om^{na}}{\Phi(a)}\;=\;V(\om^{-1});\hs
 \prod_n\:\Phi(n)\;=\;e^{-\pi i(N^2-1)/6}.\end{equation}
Writing out the factors of
 $\;\lk\,x^{N(N-1)/2}\,V(\om^{-1}\,x^{-1})\,\rk\,V(x)\:$
and extracting several powers of $\om$, we obtain
\beq     \ds V(\omega^{-1}x^{-1})\;=\;
\frac{y^{N(N-1)}}{x^{N(N-1)/2}}\;\frac{e^{i\pi(N-1)(N-2)/6}}{V(x)}\,.
\end{equation}
We can express the terms involving $p_1$ and $p_2$ using $\:V(x)$:
\[ \fl \prod_n W_{p_2}(n)\;=\;\frac{V(x_2)}{y_2^{N(N-1)/2}}\,;
\hs \prod_n\frac{1}{W_{Op_1}}\;=\;\frac{V(x_1)}{y_1^{N(N-1)/2}}\;\;
         e^{-i\pi(N^2-1)/6}.\]
In the ratio $\;V(x)/V(\om\:x)\;$ many terms cancel, leading to
\beq  \frac{V(x)}{V(\om\,x)}\;=\;\left(\frac{y}{1\,-\om\,x}\right)^{\!N}.
\end{equation}
This can be used to compute the $p_4$-term in (\ref{det-R-prel}): We define
\[ \fl F_{p=(x,y)}\;=\;\prod_n\sum_b\,\frac{\om^{n\,b}}{W_p(b)},\]
which, using
$\;W_{(\om\,x,\,y)}(b)\:=\:(1-\om\,x)\:y^{-1}\:W_{(x,y)}(b+1)\,,\;$
is seen to satisfy
\[ \fl \frac{F_{(\om\,x,\,y)}}{F_{(x,\,y)}}\;=\;
  \om^{N(N-1)/2}\lk\frac{y}{1-\om\,x}\rk^N\!\!,\hq F_{q_1}\;=\;1,
  \hq F_{q_\infty}\;=\;V^*(\om^{-1}),\hq F_{q_0}\;=\;0.\]
where $\;q_0,\;q_\infty\;$ and $q_1\;$ are the three special Fermat points
introduced in (\ref{spcq}). This is solved by
\[ \fl F_{(x,y)}\;=\;(\om\,x)^{N(N-1)/2}\frac{V(\om^{-1})}{V(x)}\,.\]
Finally, we consider the $p_3$ term: Defining
\[ \fl G_{p=(x,y)}\;=\;\prod_n\sum_a\,\frac{\om^{na}}{\Phi(a)\:W_p(a)},\]
which satisfies 
\[\fl \frac{G_{(\om\,x,\,y)}}{G_{(x,\,y)}}\;=\;\lk\frac{y}{1-\om\,x}\rk^N.\]
Using $\;\:G_{q_1}\:=\:1,\;\;G_{q_\infty}\;=\;0,\;\;
  G_{q_0}\;=\;V(\om^{-1}),$ we get $\:\;G_p\;=\;V(\om^{-1})/V(x)\,.$\\
Inserting these results into (\ref{det-R-prel}),
our final expression for $\:\det\;\R_{1,2,3}$ is:
\beq\label{det-R-fin}
\det\:\R\;=\;N^{N^3}\:\lk\lk\frac{x_4}{y_1\,y_2}\rk^{N(N-1)/2}
\frac{V(x_1)V(x_2)}{V(x_3)V(x_4)} \rk^{\!N^2}.\end{equation}
The relation $\;x_1x_2=\om\,x_3x_4\;$ has not yet been used and is
still to be imposed here.
For $\:N=2\;$ eq.(\ref{det-R-fin}) gives
\[\fl \det\:\R\;=\;2^8\lk\frac{x_4}{y_1\,y_2}\;
       \frac{(1-x_1)(1-x_2)}{(1-x_3)(1-x_4)}\rk^4\;=
       \;\lk \,4\,x_4\,\frac{y_1\,y_2}{y_3^2\,y_4^2}\;\frac{(1+x_3)(1+x_4)}
       {(1+x_1)(1+x_2)}\:\rk^4\,,\]
in agreement with (\ref{VTM}). Observe that, despite the quite similar
appearance of $W_{p_3}$ and $W_{p_4}$ in (\ref{R-N-matrix}), different
phases make (\ref{det-R-fin}) unsymmetrical between $x_3$ and $x_4$,
compare (\ref{VTM}).

\section{Parameterization of the Fermat points}

Now the most important step follows: the parameterization of the
Fermat points for each of the eight $R$-matrices. Recall that $x_4^{(j)}$
is determined by the other three $x_i^{(j)}$ due to (\ref{consg}).

Writing repeatedly the parameterization (\ref{x-uw}),(\ref{y-uw}) for the
eight $R$-matrices, then applying repeatedly the functional
mappings as it is written in (\ref{mte-general}), we get  for the
arguments appearing in (\ref{mte-elements}):
\bea && 
 x_1^{(j)}=\frac{1}{\sqrt{\om}}\:\:\XB_{j1};\hs
x_2^{(j)}=\frac{1}{\sqrt{\om}}\:\:\XB_{j2};\hs
x_3^{(j)}=\frac{1}{\om}\:\:\XB_{j3};\hs
x_4^{(j)}=\frac{x_1^{(j)}x_2^{(j)}}{\om\,x_3^{(j)}};\hs\ny\\ 
\fl &&
y_{31}^{(j)}\:=\:\YB_{j1};\hs
y_{41}^{(j)}\:=\:\frac{1}{\sqrt{\om}}\;\YB_{j2};\hs
y_{32}^{(j)}\:=\:\YB_{j3};\hs\mbox{where}\hq y_{ik}^{(j)}\;\equiv\;
\frac{y_i^{(j)}}{y_k^{(j)}}.\label{ymx}\eea
where (not writing $\;\XB_{j,4}\,$):
\beq \label{XYsys}\XB_{jk}=\!\lk\!\!\begin{array}{ccc}
\ds\frac{u_2^{(1)}}{\ka_1u_1^{(1)}}&\ds\frac{\ka_2u_2^{(2)}}{u_1^{(2)}}&
\ds\frac{u_2^{(2)}}{u_1^{(1)}}\\
\ds\frac{u_4^{(1)}}{\ka_1u_1^{(2)}}&\ds\frac{\ka_4u_4^{(3)}}{u_1^{(5)}}&
\ds\frac{u_4^{(3)}}{u_1^{(2)}}\\
\ds\frac{u_4^{(3)}}{\ka_2u_2^{(2)}}&\ds\frac{\ka_4u_4^{(5)}}{u_2^{(5)}}&
\ds\frac{u_4^{(5)}}{u_2^{(2)}}\\
\ds\frac{u_5^{(3)}}{\ka_3u_3^{(2)}}&\ds\frac{\ka_5u_5^{(5)}}{u_3^{(5)}}&
\ds\frac{u_5^{(5)}}{u_3^{(2)}}\\
\ds\frac{u_2^{(7)}}{\ka_1u_1^{(6)}}&\ds\frac{\ka_2u_2^{(5)}}{u_1^{(5)}}&
\ds\frac{u_2^{(5)}}{u_1^{(6)}}\\
\ds\frac{u_4^{(7)}}{\ka_1^{(1)}u_1^{(1)}}&
       \ds\frac{\ka_4u_4^{(5)}}{u_1^{(6)}}&
\ds\frac{u_4^{(5)}}{u_1^{(1)}}\\
\ds\frac{u_4^{(1)}}{\ka_2u_2^{(1)}}&\ds\frac{\ka_4u_4^{(7)}}{u_2^{(7)}}&
\ds\frac{u_4^{(7)}}{u_2^{(1)}}\\
\ds\frac{u_5^{(1)}}{\ka_3u_3^{(1)}}&\ds\frac{\ka_5u_5^{(8)}}{u_3^{(8)}}&
\ds\frac{u_5^{(8)}}{u_3^{(1)}}\end{array}\!\!\rk;\hs\hx
\YB_{jk}=\!\lk\!\!\begin{array}{ccc}
\ds\frac{\ka_1w_1^{(1)}}{u_3^{(2)}}&\ds\frac{\ka_3w_3^{(1)}}{w_2^{(1)}}&
\ds\frac{w_2^{(2)}}{w_3^{(1)}}\\
\ds\frac{\ka_1w_1^{(2)}}{u_5^{(3)}}&\ds\frac{\ka_5w_5^{(1)}}{w_4^{(1)}}&
\ds\frac{w_4^{(3)}}{w_5^{(1)}}\\
\ds\frac{\ka_2w_2^{(2)}}{u_6^{(4)}}&\ds\frac{\ka_6w_6^{(1)}}{w_4^{(3)}}&
\ds\frac{w_4^{(5)}}{w_6^{(1)}}\\
\ds\frac{\ka_3w_3^{(2)}}{u_6^{(5)}}&\ds\frac{\ka_6w_6^{(4)}}{w_5^{(3)}}&
\ds\frac{w_5^{(5)}}{w_6^{(4)}}\\
\ds\frac{\ka_1w_1^{(6)}}{u_3^{(5)}}&\ds\frac{\ka_3w_3^{(8)}}{w_2^{(7)}}&
\ds\frac{w_2^{(5)}}{w_3^{(8)}}\\
\ds\frac{\ka_1w_1^{(1)}}{u_5^{(5)}}&\ds\frac{\ka_5w_5^{(8)}}{w_4^{(7)}}&
\ds\frac{w_4^{(5)}}{w_5^{(8)}}\\
\ds\frac{\ka_2w_2^{(1)}}{u_6^{(5)}}&\ds\frac{\ka_6w_6^{(8)}}{w_4^{(1)}}&
\ds\frac{w_4^{(7)}}{w_6^{(8)}}\\
\ds\frac{\ka_3w_3^{(1)}}{u_6^{(8)}}&\ds\frac{\ka_6w_6^{(1)}}{w_5^{(1)}}&
\ds\frac{w_5^{(8)}}{w_6^{(1)}}
\end{array}\!\!\rk\!\!.\label{mup}\end{equation}
Here for any $\,f\,$ it is implied that
\beq\label{uw-primes} \ds
f^{(2)}=\Ropf_{1,2,3}\circ f^{(1)};\;\hq
f^{(3)}=\Ropf_{1,2,3}\Ropf_{1,4,5}\circ f^{(1)};\hq
f^{(4)}=\Ropf_{1,2,3}\Ropf_{1,4,5}\Ropf_{2,5,6}\circ f^{(1)},
\end{equation}
and
\beq\label{uw-T} \ds
f^{(5)}\;=
\Ropf_{1,2,3}\Ropf_{1,4,5}\Ropf_{2,4,6}\Ropf_{3,5,6}\circ f^{(1)}\;.
\end{equation}
For the right hand side of (\ref{mte-elements}) we define for any $f$
\beq\label{uw-stars} \ds
f^{(8)}=\Ropf_{3,5,6}\circ f^{(1)};\hq 
f^{(7)}=\Ropf_{3,5,6}\Ropf_{2,4,6}\circ f^{(1)};\hq
f^{(6)}=\Ropf_{3,5,6}\Ropf_{2,4,6}\Ropf_{1,4,5}\circ f^{(1)}\:.
\end{equation}
Due to the validity of the functional tetrahedron equation the four times
transformed function
$\;\:f^{(\bar{5})}\;=\Ropf_{3,5,6}\Ropf_{2,4,6}\Ropf_{1,4,5}\Ropf_{1,2,3}
 \circ f^{(1)}\;\:$
coincides with (\ref{uw-T}).\\[2mm]
Observe that e.g.
\[\fl u_1^{(3)}=u_1^{(4)}=u_1^{(5)};\hq 
u_2^{(2)}=u_2^{(3)};\hq u_2^{(4)}=u_2^{(5)};\hq
u_3^{(2)}=u_3^{(3)}=u_3^{(4)};\hq u_4^{(8)}=u_4^{(1)}=u_4^{(2)};\] 
\beq u_4^{(4)}=u_4^{(5)}=u_4^{(6)};\hq u_5^{(1)}=u_5^{(2)};
\hq u_5^{(3)}=u_5^{(4)};\hq u_5^{(5)}=u_5^{(6)};
\hq u_6^{(1)}=u_6^{(2)}=u_6^{(3)}.\label{uuww}\end{equation}
The MTE's leave the following four ``centers'' invariant, i.e. for 
$\;j=1,\ldots,8\;$ we have:
\beq  \cc_1=u_4^{(j)}\,u_5^{(j)}\,u_6^{(j)};\hq 
\cc_2=\frac{u_2^{(j)}\:u_3^{(j)}}{w_6^{(j)}};\hq
\cc_3=\frac{w_3^{(j)}\,w_5^{(j)}}{u_1^{(j)}};\hq
\cc_4=w_1^{(j)}\,w_2^{(j)}\,w_4^{(j)}.\label{cemte}\end{equation}
E.g. $\cc_1'=\cc_1$ because variables with indices 4,5,6 are not transformed
by $\Ropf_{1,2,3}$, then $\cc_1''=\cc_1$ since $u_4u_5$ is center of
$\Ropf_{1,4,5}$ and 6 does not appear in $\Ropf_{1,4,5}$. $\;u_5u_6$ is
center of $\Ropf_{2,5,6},\;$ etc.

Eqs.(\ref{ymx}), (\ref{mup}) provide a "free parameterization of the MTE" 
which generalizes the free parameterization of the single $\Rop_{1,2,3}$ 
introduced in remark \ref{rem-par}, p.\pageref{rem-par}  . 
It corresponds to the following scenario: We start with twenty-four arbitrary 
complex numbers $u_j^{(1)},\,w_j^{(1)},\kappa_j$, $j=1..8$, and apply 
repeatedly 
the functional mappings $\Rop^{(f)}$, choosing appropriately the phases of 
$u_{j}^{(k)},w_{j}^{(k)}$. So we obtain a 
parameterization of the eight $\R$-matrices obeying the MTE.

The natural question arises: how many \emph{independent} parameters has the 
MTE as a matrix identity? As we see already from the existence of the 
centers (\ref{cemte}), some of the twenty-four parameters will occur only
in certain combinations.  

In order to determine the independent variables in a systematic way, 
in the next subsection we shall use a simple constructive 
procedure: We express the $u_j$ and $w_j$ in terms of line section ratios, 
the parameterization being designed such as to automatically conserve the 
centers of the mapping. This is in the same spirit as the introdution of 
$\tau$-functions in the theory of solitons. We shall find that as a matrix 
identity the MTE
may be parameterized by eight independent continuous parameters and
eight discrete phases common for left and right hand sides of MTE, and besides
each of left and right hand sides contains four extra independent 
discrete parameters. In particular, the couplings $\ka_j$ can all be absorbed
by a rescaling.

Note that one great advantage of the MTEs, which is not shared by 
Yang-Baxter equations, is that many different parameterizations can be found 
and can be chosen according to the particular calculations and applications 
one likes to do.

\subsection{Parameterization in terms of line section ratios}

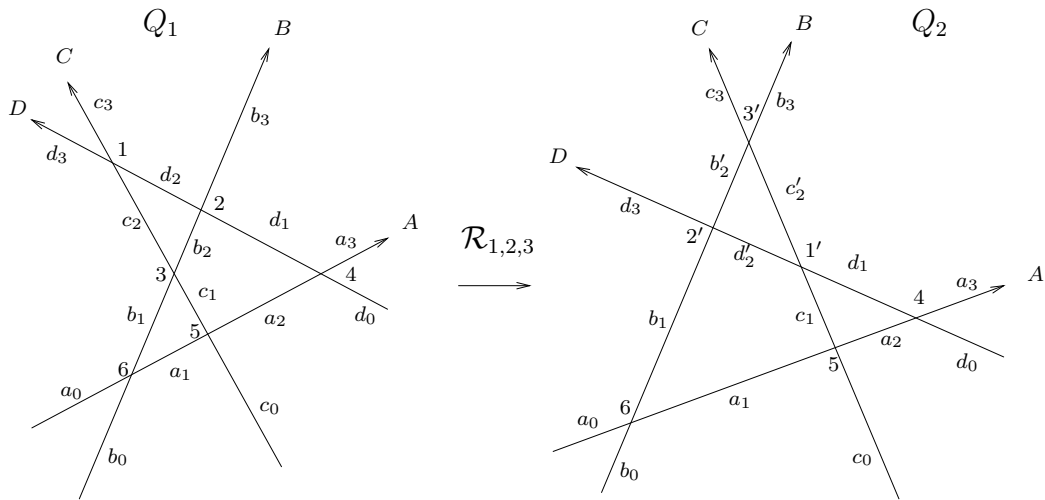
\begin{figure}[t]
\setlength{\unitlength}{0.014mm}
\begin{center}
{\renewcommand{\dashlinestretch}{30}
\begin{picture}(9813,4618)(0,-10)
\path(675,13)(2475,4288)
\path(2456.082,4165.762)(2475.000,4288.000)(2400.784,4189.045)
\path(8458,12)(6658,4287)
\path(6732.216,4188.045)(6658.000,4287.000)(6676.918,4164.762)
\path(5175,463)(9450,2038)
\path(9347.770,1968.365)(9450.000,2038.000)(9327.028,2024.666)
\path(9450,1363)(5400,3163)
\path(5521.842,3141.678)(5400.000,3163.000)(5497.473,3086.849)
\path(3600,1813)(225,3613)
\path(345.000,3583.000)(225.000,3613.000)(316.765,3530.059)
\path(2595,320)(570,3965)
\path(654.502,3874.670)(570.000,3965.000)(602.052,3845.532)
\path(225,688)(3600,2488)
\path(3508.235,2405.059)(3600.000,2488.000)(3480.000,2458.000)
\path(4275,2038)(4950,2038)
\path(4830.000,2008.000)(4950.000,2038.000)(4830.000,2068.000)
\path(5632,74)(7432,4349)
\path(7413.082,4226.762)(7432.000,4349.000)(7357.784,4250.045)
\put(4300,2308){\makebox(0,0)[lb]{${\mathcal{R}_{1,2,3}}$}}
\put(1530,1138){\makebox(0,0)[lb]{$\sk{a_1}$}}
\put(3285,1678){\makebox(0,0)[lb]{$\sk{d_0}$}}
\put(2475,2578){\makebox(0,0)[lb]{$\sk{d_1}$}}
\put(1800,1903){\makebox(0,0)[lb]{$\sk{c_1}$}}
\put(2300,3568){\makebox(0,0)[lb]{$\sk{b_3}$}}
\put(1750,2308){\makebox(0,0)[lb]{$\sk{b_2}$}}
\put(2430,1650){\makebox(0,0)[lb]{$\sk{a_2}$}}
\put(1125,1678){\makebox(0,0)[lb]{$\sk{b_1}$}}
\put(3735,2578){\makebox(0,0)[lb]{$\sk{A}$}}
\put(2520,4423){\makebox(0,0)[lb]{$\sk{B}$}}
\put( 450,4153){\makebox(0,0)[lb]{$\sk{C}$}}
\put(   0,3658){\makebox(0,0)[lb]{$\sk{D}$}}
\put( 810,3703){\makebox(0,0)[lb]{$\sk{c_3}$}}
\put( 360,3208){\makebox(0,0)[lb]{$\sk{d_3}$}}
\put(1440,3000){\makebox(0,0)[lb]{$\sk{d_2}$}}
\put(1080,2578){\makebox(0,0)[lb]{$\sk{c_2}$}}
\put(945,328){\makebox(0,0)[lb]{$\sk{b_0}$}}
\put(495,958){\makebox(0,0)[lb]{$\sk{a_0}$}}
\put(2385,823){\makebox(0,0)[lb]{$\sk{c_0}$}}
\put(3105,2398){\makebox(0,0)[lb]{$\sk{a_3}$}}
\put(8010,328){\makebox(0,0)[lb]{$\sk{c_0}$}}
\put(7965,2173){\makebox(0,0)[lb]{$\sk{d_1}$}}
\put(5805,2713){\makebox(0,0)[lb]{$\sk{d_3}$}}
\put(5130,3208){\makebox(0,0)[lb]{$\sk{D}$}}
\put(6480,4423){\makebox(0,0)[lb]{$\sk{C}$}}
\put(7470,4468){\makebox(0,0)[lb]{$\sk{B}$}}
\put(9675,2083){\makebox(0,0)[lb]{$\sk{A}$}}
\put(9000,2000){\makebox(0,0)[lb]{$\sk{a_3}$}}
\put(9000,1228){\makebox(0,0)[lb]{$\sk{d_0}$}}
\put(8280,1450){\makebox(0,0)[lb]{$\sk{a_2}$}}
\put(6840,880){\makebox(0,0)[lb]{$\sk{a_1}$}}
\put(5400,688){\makebox(0,0)[lb]{$\sk{a_0}$}}
\put(5805,193){\makebox(0,0)[lb]{$\sk{b_0}$}}
\put(7470,1678){\makebox(0,0)[lb]{$\sk{c_1}$}}
\put(6885,2218){\makebox(0,0)[lb]{$\sk{d'_2}$}}
\put(7380,2848){\makebox(0,0)[lb]{$\sk{c'_2}$}}
\put(7290,3703){\makebox(0,0)[lb]{$\sk{b_3}$}}
\put(6615,3793){\makebox(0,0)[lb]{$\sk{c_3}$}}
\put(6660,3073){\makebox(0,0)[lb]{$\sk{b'_2}$}}
\put(1035,3270){\makebox(0,0)[lb]{$\sk{1}$}}
\put(1950,2758){\makebox(0,0)[lb]{$\sk{2}$}}
\put(1395,2083){\makebox(0,0)[lb]{$\sk{3}$}}
\put(3200,2083){\makebox(0,0)[lb]{$\sk{4}$}}
\put(1720,1543){\makebox(0,0)[lb]{$\sk{5}$}}
\put(1040,1183){\makebox(0,0)[lb]{$\sk{6}$}}
\put(8595,1858){\makebox(0,0)[lb]{$\sk{4}$}}
\put(7785,1228){\makebox(0,0)[lb]{$\sk{5}$}}
\put(6435,2443){\makebox(0,0)[lb]{$\sk{2'}$}}
\put(6975,3613){\makebox(0,0)[lb]{$\sk{3'}$}}
\put(7560,2308){\makebox(0,0)[lb]{$\sk{1'}$}}
\put(5805, 823){\makebox(0,0)[lb]{$\sk{6}$}}
\put(6075,1633){\makebox(0,0)[lb]{$\sk{b_1}$}}
\put(1275,4400){\makebox(0,0)[lb]{$Q_1$}}
\put(8575,4400){\makebox(0,0)[lb]{$Q_2$}}
\end{picture}}
\caption{\footnotesize{
Parameterization of the arguments 
$u_i^{(j)},\;w_i^{(j)}$ of the rational mapping 
$\mathcal{R}_{i,j,k}^{(j)}$ in terms of line-section ratios.
}}\label{tau-par}
\end{center}
\end{figure}

We now parameterize the $u_i^{(j)},\;w_i^{(j)}$ in terms of ratios of 
parameters which can be read off from the four line graphs ("quadrilaterals") 
shown in Fig.~\ref{tet-graph}. Fig.~\ref{tau-par} gives an enlarged detail of 
Fig.~\ref{tet-graph} with labels attached, which are explained in the 
following:
\\[1mm]
In the quadrilateral $\:Q_1\:$ there are four directed lines 
$A,\:B,\:C,\:D$. The six vertex points cut the lines into four sections 
each. We denote the sections
of line $A$ by $a_0,\;a_1,\;\ldots,\;a_3$ 
(the indices increasing in the direction of the line). 
Analogously, we label the sections of lines $B,\;C,\;D\;$ by 
$b_0,\ldots,\:d_3$.  
This way for each quadrilateral $Q_j$ we have defined 16 variables.\\[1mm]
Now, to each vertex $i$ of each quadrilateral $Q_j$ we associate a pair of 
variables $u_i^{(j)},\;w_i^{(j)}$ (which determine the Fermat points in 
(\ref{ymx})), and we require these to be given in terms of the 
$\;a_0,\:a_1,\:\ldots,\:d_2,\:d_3\;$ of the 
respective quadrilateral as follows:\\[1mm]
For a $u_i^{(j)}$, take $Q_j$ and look from point $i$ in the direction of 
the arrows and select the right-pointing of the two lines. Now $u_i^{(j)}$ 
is the ratio of the variable attached
to the section {\it before} to the variable {\it after} the vertex. 
For the $w_i^{(j)}$ take the left-pointing line, and divide the variable 
after the vertex by the variable before the vertex. 
So, in the expressions for the $u_i^{(j)}$ the index of the numerator 
is one smaller than that for the denominator, the inverse is true for the 
$w_i^{(j)}$.
\\[1mm] 
Passing from one quadrilateral to the next one, corresponding to a mapping 
$\Rop$, always three of the ``internal`` lines are changed, and for 
distinction, we attach to these changed variables dashes and daggers. 
The eight ``external`` variables 
$a_0,\;a_3,\,\ldots,\:d_0,\;d_3$ are never changed by our mappings.  
Of the eight ``internal`` variables (these have the indices 1 and 2) five are
unchanged in each mapping. 
To collect these definitions, we define
\beq U_j=\left[{u_1^{(j)}}\!,\;{u_2^{(j)}}\!,\;\ldots,\;{u_5^{(j)}}\!,
   \;{u_6^{(j)}}\!,
\;{w_1^{(j)}}\!,\;{w_2^{(j)}}\!,\;\ldots,\;{w_5^{(j)}}\!,\;{w_6^{(j)}}\right].
   \label{uj}\end{equation}
Then from Figs.~\ref{tet-graph},~\ref{tau-par}~ we read off, applying 
successively the anticlockwise mappings:
\[\fl U_1=\left[\frac{c_2}{c_3},\;\frac{b_2}{b_3},
     \;\frac{b_1}{b_2},\;
\frac{a_2}{a_3},\;\frac{a_1}{a_2},\;\frac{a_0}{a_1},\;
\frac{d_3}{d_2},\;\frac{d_2}{d_1},\;\frac{c_2}{c_1},\;
\frac{d_1}{d_0},\;\frac{c_1}{c_0},\;\frac{b_1}{b_0}
        \right];\]
\[\fl U_2=\left[\frac{c_1}{c_2'},\;\frac{b_1}{{b_2}'},
     \;\frac{{b_2}'}{b_3},\;
\frac{a_2}{a_3},\;\frac{a_1}{a_2},\;\frac{a_0}{a_1},\;
\frac{{d_2}'}{d_1},\;\,\frac{d_3}{{d_2}'},\;\,\frac{c_3}{c_2'},\;
\frac{d_1}{d_0},\;\,\frac{c_1}{c_0},\;\frac{b_1}{b_0}\right];\]
\[\fl U_3=\left[\frac{c_0}{{c_1''}},\;\frac{b_1}{{b_2'}},
     \:\,\frac{{b_2'}}{b_3},\;\frac{a_1}{{a_2''}},\:\,
\frac{{a_2''}}{a_3},\:\frac{a_0}{a_1},\:\,
\frac{{d_1''}}{d_0},\:\,\frac{d_3}{{d_2}'},\:\,\frac{c_3}{c_2'},
\:\,\frac{{d_2}'}{{d_1''}},\:\,\frac{{c_2'}}{{c_1''}},\:
\frac{b_1}{b_0}\right];\]
\[\fl U_4=\left[\frac{c_0}{{c_1''}},\:\frac{b_0}{{b_1'''}},
     \:\frac{{b_2'}}{b_3},\:\frac{a_0}{{a_1'''}},\:
\frac{{a_2''}}{a_3},\:\frac{{a_1'''}}{{a_2''}},\:
\frac{{d_1''}}{d_0},\:\frac{{d_2'''}}{{d_1}''},\:
\frac{c_3}{c_2'},\:\frac{{d_3}}{{d_2'''}},\:
\frac{{c_2'}}{{c_1''}},\:\frac{{b_2}'}{{b_1}'''}\right];\]
\beq  U_5=\left[\frac{c_0}{c_1''},\:\frac{b_0}{b_1'''},
     \:\frac{b_1'''}{{b_2^T}},\:
    \frac{a_0}{a_1'''},\:\frac{a_1'''}{{a_2^T}},
    \:\frac{{a_2^T}}{a_3},\:
    \frac{d_1''}{d_0},\:\frac{d_2'''}{d_1''},
    \:\frac{{c_2^T}}{c_1''},\:
    \frac{d_3}{d_2'''},\:\frac{c_3}{{c_2^T}},
     \:\frac{b_3}{{b_2^T}}\right];\label{ufunf}\end{equation}
We write $a_2^T$ instead of $a_2''''$ and $d_1^t$ 
instead of $d_1^{\dg\dg\dg\dg}$, etc.
Transforming clockwise in Fig.~\ref{tet-graph}, we get
\[\fl U_8=\left[\frac{c_2}{c_3},\;\frac{b_2}{b_3},
     \;\frac{b_0}{b_1^\dg},\;
\frac{a_2}{a_3},\;\frac{a_0}{a_1^\dg},\;\frac{a_1^\dg}{a_2},\;
\frac{d_3}{d_2},\;\frac{d_2}{d_1},\;\frac{c_1^\dg}{c_0},\;
\frac{d_1}{d_0},\;\frac{c_2}{c_1^\dg},\;\frac{b_2}{b_1^\dg}
        \right],\]
\[\fl U_7=\left[\frac{c_2}{c_3},\;\frac{b_1^\dg}{b_2^\dgg},
     \;\frac{b_0}{b_1^\dg},\;
\frac{a_1}{a_2^\dgg},\;\frac{a_0}{a_1^\dg},\;\frac{a_2^\dgg}{a_3},\;
\frac{d_3}{d_2},\;\frac{d_1^\dgg}{d_0},\;\frac{c_1^\dg}{c_0},\;
\frac{d_2}{d_1^\dgg},\;\frac{c_2}{c_1^\dg},\;\frac{b_3}{b_2^\dgg} \right];\]
etc. Transforming clockwise two times more, we arrive
at an alternative expression for $\:U_5\:$, which for distinction we call 
$\:U_{\bar{5}}\,:$ 
\beq  U_{\bar{5}}=\left[\frac{c_0}{{c_1^t}},\:\frac{b_0}{{b_1^t}},
    \:\frac{{b_1^t}}{{b_2^\dgg}},\:
    \frac{a_0}{{a_1^\dgh}},\:\frac{{a_1^\dgh}}{{a_2^\dgg}},
    \:\frac{{a_2^\dgg}}{a_3},\:
    \frac{{d_1^t}}{d_0},\:\frac{{d_2^\dgh}}{{d_1^t}},
    \:\frac{{c_2^\dgh}}{{c_1^t}},\:
 \frac{d_3}{{d_2^\dgh}},\:\frac{c_3}{{c_2^\dgh}},
 \:\frac{b_3}{{b_2^\dgg}}\right]. \label{ubfunf} \end{equation}
We see that this particular parameterization in terms of ratios  
automatically incorporates  
the invariance of the centers of the mappings. E.g. for $\Rop^{(1)}$ 
(in our notation (\ref{uj}) $\:u_j^{(2)}$ 
is the $\:u_j'$ of (\ref{f-centers})):
\[\fl   u_2^{(2)}u_3^{(2)}=u_2^{(1)}u_3^{(1)};
\hs u_1^{(2)}/w_3^{(2)}=u_1^{(1)}/w_3^{(1)};
\hs w_1^{(2)}w_2^{(2)}=w_1^{(1)}w_2^{(1)}. \]
Also more complicated conditions are fulfilled automatically, e.g.
\[ \fl u_1^{(4)}w_3^{(1)}w_5^{(1)}\;=\;u_1^{(1)}w_3^{(2)}w_5^{(3)}.\]
Observe that because in the mapping from $Q_1$ to $Q_2$ the line $A$
keeps the same intersection points, there is no $a_i'$ in the 
formulae for the $U_j$, similarly, there are no $\;b_i'',\;c_i''',\;d_i^T,\;
d_i^\dg,\;c_i^{\dgg},$ $b_i^{\dgh},\;a_i^t.$ So, altogether, there
are 24 primed or daggered variables appearing. However, some coincide 
as we will see. From (\ref{mapping-N}) and (\ref{lambda-N}) we easily 
obtain the following relations, which allow to express their $N$-$th$ powers 
recursively in terms of the $N$-$th$ powers of the unprimed variables:
\bea  
{b_2}'^Nc_2^Nd_2^N&=&b_1^Nc_3^Nd_2^N+b_2^Nc_3^Nd_3^N
+\ka_1^Nb_3^Nc_2^Nd_3^N;\ny\\
\fl \ka_2^Nb_2^Nc_2'^Nd_2^N&=&\ka_1^Nb_3^Nc_1^Nd_2^N+\ka_3^Nb_2^Nc_3^Nd_1^N
+\ka_1^N\ka_3^Nb_3^Nc_2^Nd_1^N;\ny\\
\fl b_2^Nc_2^N{d_2}'^N&=&b_2^Nc_1^Nd_3^N+b_1^Nc_1^Nd_2^N
+\ka_3^Nb_1^Nc_2^Nd_1^N;\ny\\[1mm]
\fl a_2''^Nc_1^Nd_1^N&=&a_1^Nc_2'^Nd_1^N+a_2^Nc_2'^N{d_2}'^N
+\ka_1^Na_3^Nc_1^N{d_2}'^N;\ny\\
\fl \ka_4^Na_2^Nc_1''^Nd_1^N&=&\ka_1^Na_3^Nc_0^Nd_1^N+\ka_5^Na_2^Nc_2'^Nd_0^N
+\ka_1^N\ka_5^Na_3^Nc_1^Nd_0^N;\ny\\
\fl a_2^Nc_1^Nd_1''^N&=&a_2^Nc_0^N{d_2}'^N+a_1^Nc_0^Nd_1^N
+\ka_5^Na_1^Nc_1^Nd_0^N;\ny\\[1mm]
\fl a_1'''^Nb_1^N{d_2}'^N&=&a_0^N{b_2}'^N{d_2}'^N+d_3^Na_1^N{b_2}'^N
+\ka_2^Nd_3^Nb_1^Na_2''^N;\ny\\
\fl \ka_4^Na_1^Nb_1'''^N{d_2}'^N&=&\ka_2^Nb_0^Na_2''^N{d_2}'^N
+\ka_6^Na_1^N{b_2}'^Nd_1''^N
 +\ka_2^N\ka_6^Na_2''^Nb_1^Nd_1''^N;\ny\\
\fl a_1^Nb_1^N{d_2}'''^N&=&b_0^Nd_3^Na_1^N+a_0^Nb_0^N{d_2}'^N
+\ka_6^Na_0^Nb_1^Nd_1''^N;\ny\\[1mm]
\fl {a_2^T}^N{b_2}'^Nc_2'^N&=&b_3^Na_1'''^Nc_2'^N+b_3^Nc_3^Na_2''^N
+\ka_3^Na_3^Nc_3^N{b_2}'^N;\ny\\
\fl \ka_5^Na_2''^N{b_2^T}^Nc_2'^N&=&\ka_3^Na_3^Nb_1'''^Nc_2'^N
+\ka_6^Nb_3^Na_2''^Nc_1''^N
  +\ka_3^N\ka_6^Na_3^N{b_2}'^Nc_1''^N;\ny\\
\fl a_2''^N{b_2}'^Nc_2^T&=&c_3^Na_2''^Nb_1'''^N+a_1'''^Nb_1'''^Nc_2'^N
+\ka_6^Na_1'''^N{b_2}'^Nc_1''^N;
\ny \\[1mm]
\fl {a_1^\dg}^N b_1^Nc_1^N&=&a_0^Nb_2^Nc_1^N+a_1^Nb_2^Nc_2^N
+\ka_3^Na_2^Nb_1^Nc_2^N;\ny\\
\fl \ka_5^Na_1^N{b_1^\dg}^N c_1^N&=&
\ka_3^Na_2^Nb_0^Nc_1^N+\ka_6^Na_1^Nb_2^Nc_0^N
+\ka_3^N\ka_6^Na_2^Nb_1^Nc_0^N;\;\ny\\
\fl a_1^Nb_1^N{c_1^\dg}^N&=&a_1^Nb_0^Nc_2^N+a_0^Nb_0^Nc_1^N
+\ka_6^Na_0^Nb_1^Nc_0^N;\ny \\[1mm]
\fl  a_2^Nb_2^N{d_1^{\dgg}}^N&=&
             a_2^N{b_1^\dg}^N d_2^N+{a_1^\dg}^N {b_1^\dg}^N d_1^N
            +\ka_6^N{a_1^\dg}^N b_2^Nd_0^N;\ny\\
\fl {a_2^{\dgg}}^Nb_2^Nd_1^N&=&b_3^N{a_1^\dg}^N d_1^N+b_3^Na_2^Nd_2^N
+\ka_2^Na_3^Nb_2^Nd_2^N;\ny\\
\fl \ka_4^Na_2^N{b_2^{\dgg}}^Nd_1^N&=&
\ka_2^Na_3^N{b_1^\dg}^N d_1^N+\ka_6^Nb_3^Na_2^Nd_0^N
+\ka_2^N\ka_6^Na_3^Nb_2^Nd_0^N;
          \ny\\[1mm]
\fl {a_1^{\dgh}}^Nc_2^Nd_2^N&=&a_0^Nc_3^Nd_2^N+c_3^Nd_3^Na_1'^N
+\ka_1^Nd_3^Na_2''^Nc_2^N;\ny\\
\fl \ka_4^N{a_1^\dg}^N {c_2^{\dgh}}^Nd_2^N&=&
\ka_1^Na_2^{\dgg}{c_1^\dg}^N d_2^N+\ka_5^Nc_3^N{a_1^\dg}^N {d_1^{\dgg}}^N
  +\ka_1^N\ka_5^N{a_2^{\dgg}}^Nc_2^N{d_1^{\dgg}}^N;\ny\\
\fl {a_1^\dg}^N c_2^N{d_2^{\dgh}}^N&=&
d_3^N{a_1^\dg}^N {c_1^\dg}^N +a_0^N{c_1^\dg}^N d_2^N
+\ka_5^Na_0^Nc_2^N{d_1^{\dgg}}^N;\ny\\[1mm]
\fl {b_1^t}^N{c_1^\dg}^N {d_1^{\dgg}}^N&=&b_0^N{c_2^{\dgh}}^N{d_1^{\dgg}}^N
+{b_1^\dg}^N {c_2^{\dgh}}^N{d_2^{\dgh}}^N
             +\ka_1^N{b_2^{\dgg}}^N{c_1^\dg}^N {d_2^{\dgh}}^N;\ny\\
\fl \ka_2^N{b_1^\dg}^N {c_1^t}^N{d_1^{\dgg}}^N&=&
\ka_1^Nc_0^N{b_2^{\dgg}}^N{d_1^{\dgg}}^N
          +\ka_3^Nd_0^N{b_1^\dg}^N {c_2^{\dgh}}^N
          +\ka_1^N\ka_3^Nd_0^N{b_2^{\dgg}}^N{c_1^\dg}^N;\ny\\
\fl {b_1^\dg}^N {c_1^\dg}^N {d_1^{\,t}}^N&=&
c_0^N{b_1^\dg}^N {d_2^{\dgh}}^N+b_0^Nc_0^N{d_1^{\dgg}}^N
+\ka_3^Nb_0^Nd_0^N{c_1^\dg}^N.
\label{deri}\eea  
Using these relations one verifies by straightforward
substitution in (\ref{ufunf}) and (\ref{ubfunf}) that 
\beq U_5\;=\;U_{\bar{5}},\label{fcz}\end{equation} 
i.e. that the functional tetrahedron equations are satisfied. Eq.(\ref{fcz}) 
tells us that there are eight direct relations between the dashed and daggered
variables, e.g. $c_1''=c_1^t$, etc. so that in fact due to the functional 
tetrahedron equations the last eight equations of (\ref{deri}) are superfluous. 
This will be important for the discussion of
the freedom of phase choices when taking $Nth$ roots.  \\[2mm]
Written in terms of our parameters $a_i,\;b_i,$ etc. the arguments of
the $\Rop^{(j)}\;\;(j=1,\ldots,8)$, see (\ref{ymx}), are
\bea x_1^{(j)}=\frac{1}{\sqrt{\om}}\:\XM_{j1};&&\hs
x_2^{(j)}=\frac{1}{\sqrt{\om}}\:\XM_{j2};\hs
x_3^{(j)}=\frac{1}{\om}\:\XM_{j3};\hs
x_4^{(j)}=\frac{x_1^{(j)}x_2^{(j)}}{\om\,x_3^{(j)}};\ny\\
\fl y_{31}^{(j)}\:=\:\YM_{j1};&&\hs\hq
y_{41}^{(j)}\:=\:\frac{1}{\sqrt{\om}}\:\YM_{j2};\hs\hs
y_{32}^{(j)}\:=\:\YM_{j3},\label{xmx}\eea
where (again we do not write out $\;x_4^{(j)}\,$)
\beq \XM_{jk}=\!\lk\!\!\begin{array}{ccc}
\ds\frac{b_2c_3}{\ka_1b_3c_2}&\ds\frac{\ka_2b_1c_2'}{{b_2}'c_1}&
\ds\frac{b_1c_3}{{b_2}'c_2}\\
\ds\frac{a_2c_2'}{\ka_1a_3c_1}&\ds\frac{\ka_4a_1c_1''}{a_2''c_0}&
\ds\frac{a_1c_2'}{a_2''c_1}\\
\ds\frac{a_1{b_2}'}{\ka_2a_2''b_1}&\ds\frac{\ka_4a_0b_1'''}{b_0a_1'''}&
\ds\frac{a_0{b_2}'}{a_1'''b_1}\\
\ds\frac{a_2''b_3}{\ka_3{b_2}'a_3}&\ds\frac{\ka_5a_1'''b_2^T}{b_1'''a_2^T}&
\ds\frac{a_1'''b_3}{a_2^T{b_2}'}\\
\ds\frac{b_1^\dg c_2^{\dgh}}{\ka_1b_2^{\dgg}c_1^\dg}&
\ds\frac{\ka_2b_0c_1^t}{b_1^tc_0}&\ds
     \frac{b_0c_2^{\dgh}}{b_1^tc_1^\dg}\\
\ds\frac{a_1^\dg c_3}{\ka_1a_2^{\dgg}c_2}&
\ds\frac{\ka_4a_0c_2^{\dgh}}{c_1^\dg a_1^{\dgh}}&\ds
     \frac{a_0c_3}{a_1^{\dgh}c_2}\ny\\
\ds\frac{a_2b_3}{\ka_2a_3b_2}&
\ds\frac{\ka_4a_1^\dg b_2^{\dgg}}{a_2^{\dgg}b_1^\dg}&\ds
     \frac{a_1^\dg b_3}{a_2^{\dgg}b_2}\\
\ds\frac{a_1b_2}{\ka_3a_2b_1}&\ds\frac{\ka_5a_0b_1^\dg}{b_0a_1^\dg}&\ds
     \frac{a_0b_2}{a_1^\dg b_1}\end{array}\!\!\rk\!;\hx
\YM_{jk}=\!\lk\!\!\!\begin{array}{ccc}
\ds\frac{\ka_1d_3b_3}{d_2{b_2}'}&\ds\frac{\ka_3c_2d_1}{c_1d_2}&
\ds\frac{d_3c_1}{{d_2}'c_2}\\
\ds\frac{\ka_1{d_2}\!'a_3}{d_1a_2''}&\ds\frac{\ka_5c_1d_0}{c_0d_1}&
\ds\frac{{d_2}'c_0}{{d_1}''c_1}\\
\ds\frac{\ka_2d_3a_2''}{{d_2}'a_1'''}&\ds\frac{\ka_6b_1{d_1}''}{b_0{d_2}'}&
\ds\frac{d_3b_0}{{d_2}'''b_1}\\
\ds\frac{\ka_3c_3a_3}{c_2'a_2^T}&\ds\frac{\ka_6{b_2}'c_1''}{{b_1}'''c_2'}&
\ds\frac{c_3{b_1}'''}{c_2^T{b_2}'}\\
\ds\frac{\ka_1d_2^{\dgh}b_2^{\dgg}}{d_1^{\dgg}b_1^t}&
\ds\frac{\ka_3c_1^\dg d_0}{c_0d_1^{\dgg}}&
         \ds\frac{d_2^{\dgh}c_0}{d_1^tc_1^\dg}\\
\ds\frac{\ka_1d_3a_2^{\dgg}}{d_2a_1^{\dgh}}&
\ds\frac{\ka_5c_2d_1^{\dgg}}{c_1^\dg d_2}&\ds\frac{d_3c_1^\dg}
  {d_2^{\dgh}c_2}\\
\ds\frac{\ka_2d_2a_3}{d_1a_2^{\dgg}}&\ds\frac{\ka_6b_2d_0}{b_1^\dg d_1}&
\ds\frac{d_2b_1^\dg}{d_1^{\dgg}b_2}\\
\ds\frac{\ka_3c_2a_2}{c_1a_1^\dg}&\ds\frac{\ka_6b_1c_0}{b_0c_1}&
\ds\frac{c_2b_0}{c_1^\dg b_1}
\end{array}\!\!\rk\!\!.\label{xy}\end{equation}
To check the validity of all the Fermat relations
$\;{x_i^{(j)}}^N\,+{y_i^{(j)}}^N\,=\,1$ requires using the 
transformation equations (\ref{deri}).
The relations (\ref{deri}) involve the $Nth$ powers of the variables, so
to use them to get the Fermat coordinates, we have to take $Nth$ roots,
which entails discrete phase choices. 
The centers (\ref{cemte}) are just ratios of the external variables:
\beq \cc_1=\frac{a_0}{a_3};\hs\cc_2=\frac{b_0}{b_3};
\hs\cc_3=\frac{c_3}{c_0};\hs\cc_4=\frac{d_3}{d_0}.\end{equation}

The external variables $\;a_0,\;a_3,\;\ldots,\;d_0,\;d_3\;$ are irrelevant 
and serve mainly to express all quantities in terms of ratios.
We may choose them simply all to be unity. The "coupling constants" $\ka_j$ 
may all be eliminated by re-scaling the eight relevant variables 
$\;a_1,\;a_2,\;\ldots,\;d_1,\;d_2\;$ as follows:
\beq\begin{tabular}{llll}
    $a_1\:=\:\ds\frac{\ka_1\ka_2}{\ka_5}\:\overline{a_1}; $&
$\hq a_2\:=\:\ds\frac{\ka_1\ka_6}{\ka_3}\:\overline{a_2};\hq$ &
    $b_1\:=\:\ds\frac{\ka_1}{\ka_3\ka_5}\:\overline{b_1}; $&$
\hq  b_2\:=\:\ds\frac{\ka_1\ka_6}{\ka_2\ka_3}\:\overline{b_2};$\\[3mm]
    $c_1\:=\:\ds\frac{\ka_1\ka_6}{\ka_3\ka_5}\:\overline{c_1}; $&$
\hq  c_2\:=\:\ds\frac{\ka_6}{\ka_2\ka_3}\:\overline{c_2};\hq$ &
    $d_1\:=\:\ds\frac{\ka_1\ka_6}{\ka_3}\:\overline{d_1}; $&$ 
\hq  d_2\:=\:\ds\frac{\ka_5\ka_6}{\ka_2}\:\overline{d_2}.$  
\end{tabular}   \label{resca}\end{equation}
This entails a corresponding re-scaling of the dashed and daggered 
variables, e.g. 
\beq\begin{tabular}{llll}
    ${b_2}'\:=\:\ds\frac{\ka_1\ka_2}{\ka_5\ka_6}\:\overline{{b_1}'}; $&
$\hq {c_2}'\:=\:\ds\frac{\ka_1}{\ka_5}\:\overline{{c_2}'};\hq$ &
    ${d_2}'\:=\:\ds\frac{\ka_1\ka_2}{\ka_5}\:\overline{{d_2}'}; $&$
\hq  a_2''\:=\:\ds\frac{\ka_1\ka_2\ka_3}{\ka_5\ka_6}\:\overline{a_2''};$
\\[3mm]
    $c_1''\:=\:\ds\frac{\ka_3}{\ka_4\ka_6}\:\overline{c_1''}; $&$
\hq  d_1''\:=\:\ds\frac{\ka_2\ka_3}{\ka_6}\:\overline{d_1''};\hq$ &
    $a_1'''\:=\:\ds\frac{\ka_2\ka_3}{\ka_6}\:\overline{a_1'''}; $&$ 
\hq  b_1'''\:=\:\ds\frac{\ka_2\ka_3}{\ka_4\ka_6}\:\overline{b_1'''};$  
\\[3mm]
    $d_2'''\:=\:\ds\frac{\ka_3\ka_5}{\ka_1}\:\overline{d_2'''}; $&$
\hq  a_2^t\:=\:\ds\frac{\ka_3\ka_5}{\ka_1}\:\overline{a_2^t};\hq$ &
    $b_2^t\:=\:\ds\frac{\ka_3}{\ka_1\ka_4}\:\overline{b_2^t}; $&$ 
\hq  c_2^t\:=\:\ds\frac{\ka_3\ka_5}{\ka_1\ka_4}\:\overline{c_2^t};$  
\\[3mm]
    $a_1^\dg\:=\:\ds\frac{\ka_5\ka_6}{\ka_2}\:\overline{a_1^\dg}; $&$
\hq  b_1^\dg\:=\:\ds\frac{\ka_6}{\ka_2}\:\overline{b_1^\dg};\hq$ &
    $c_1^\dg\:=\:\ds\frac{\ka_5\ka_6}{\ka_1\ka_2}\:\overline{c_1^\dg}; $&$ 
\hq  \ldots$  
\end{tabular}  \label{rescb}\end{equation}
So we can simplify eqs.(\ref{deri}) and (\ref{xy}) by taking 
$\:a_0=a_3=b_0=\ldots=d_3=1 \:$ and $\;\ka_1=\ka_2=\ldots=\ka_6=1\:$ and 
replacing the relevant variables by their overlined counterparts.

\subsection{The choice of discrete phases.}\label{disph}
As we already mentioned,
apart from the eight continuous parameters just discussed, the left hand 
and right hand sides of the MTE depend on phases choices arising from taking
$Nth$ roots. We investigate how many independent choices can be made and 
whether these affect the MTEs. 

From (\ref{fcz}) and (\ref{ufunf}),~(\ref{ubfunf}) we see that the left 
hand side (LHS) and 
right hand side (RHS) of the MTE have 8 arbitrary {\it common} phases of 
\[\fl a_2^T=a_2^{\dg\dg}\,,\;b_2^T=b_2^{\dg\dg}\,,\; 
c_2^T=c_2^{\dg\dg\dg}\,,\;d_2'''=d_2^{\dg\dg\dg}\,,\;\;\;
a_1'''=a_1^{\dg\dg\dg}\,,\; b_1'''=b_1^t\,,\;c_1''=c_1^t\,,\;d_1''=d_1^t.\]
These phases correspond to the phases of $u_1^{(5)}$ etc.
Furthermore, the LHS of the MTE contains 4 internal phases of 
\[\fl c_2',\;{b_2}',\;{d_2}',\;a_2''\]
while the RHS contains 4 internal phases of
\[\fl a_1^\dg,\;c_1^\dg,\;b_1^\dg,\;d_1^{\dg\dg}\;.\]
In which way does the LHS depend on its internal phases?
Consider e.g. the shift
\[\fl c_2'\;\mapsto\; q^{-1}c_2'\,.\]
According to (\ref{xy}) this shift changes the following Fermat coordinates:
\[\fl x_2^{(1)}\;\mapsto\; q^{-1}\,x_2^{(1)},\hs 
x_4^{(1)}\;\mapsto\; q^{-1}x_4^{(1)}\;,\]
\[\fl  x_1^{(2)}\;\mapsto\; q^{-1}\,x_1^{(2)},\hs 
x_3^{(2)}\;\mapsto\; q^{-1}x_3^{(2)}\;,\hs
y_1^{(4)}\;\mapsto\; q\, y_1^{(4)}\,.\]
Now note that for $\;q\:=\:\om$
\[\fl W_{(q^{-1}x,y)}(n)=\frac{y}{1-x}W_{(x,y)}(n-1)\;;\hs
W_{(x,qy)}(n)=q^nW_{(x,y)}(n)\;.\]
Therefore our shift produces \[\fl \begin{array}{l}
\langle i_1,i_2,i_3|R^{(1)}|j_1,j_2,j_3\rangle \;\;\mapsto\;\;
\ds\frac{y_2^{(1)}}{y_4^{(1)}}\frac{1-x_4^{(1)}}{1-x_2^{(1)}}\,
\langle i_1,i_2,i_3|R^{(1)}|j_1+1,j_2,j_3\rangle\: q^{-j_3}\;,\\
\\
\langle j_1,i_4,i_5|R^{(2)}|k_1,j_4,j_5\rangle\;\;\mapsto\;\;
\ds\frac{y_1^{(2)}}{y_3^{(2)}}\frac{1-x_3^{(2)}}{1-x_1^{(2)}}\,
\langle j_1+1,i_4,i_5|R^{(2)}|k_1,j_4,j_5\rangle\: q^{j_5}\;,\\
\\
\langle j_3,j_5,j_6|R^{(4)}|k_3,k_5,k_6\rangle\;\;\mapsto\;\;
q^{j_3-j_5}\,
\langle j_3,j_5,j_6|R^{(4)}|k_3,k_5,k_6\rangle\;.
\end{array}\]
We see that the change considered produces a simple scalar factor
and does not change the matrix structure of the LHS of the MTE:
\[\fl LHS(q^{-1}c_2')\;=\;
\frac{y_2^{(1)}}{y_4^{(1)}}\;
\frac{1-x_4^{(1)}}{1-x_2^{(1)}}\;
\frac{y_1^{(2)}}{y_3^{(2)}}\;
\frac{1-x_3^{(2)}}{1-x_1^{(2)}}\;\;LHS(c_2')\]
In this way we may convince ourselves:
\begin{itemize}
\item A change of the phases of any of the eight internal 
variables produces only extra scalar factors for the LHS or RHS 
of the MTE, while the external matrix structure does not change.
\item A change of the phases for any of the eight external 
variables produces extra scalar factors both for the LHS and RHS 
of the MTE, and besides it produces a change of the external 
matrix structure: a shift of the indices $i_1,...,k_6$ and 
some multipliers $q^{\pm i_1},...,q^{\pm k_6}$. However, this 
change is the same for the LHS and RHS of the MTE.
\end{itemize}

\section{Explicit form of the MTE for $N=2$}

For $N=2$, using combined indices
$\;i=1+i_1+2 i_2+4 i_3;\;\;k=1+k_1+2 k_2+4 k_3\,,\:$ we can give
$\lk R\rk_{i1,i2,i3}^{k1,k2,k3}$ explicitly in a simple matrix form:
We define
\bea Y_k&=&\frac{y_k}{1\,+\,x_k}\;=\;\sqrt{\frac{1-x_k}{1+x_k}}
  \hs\mbox{for}\hs k=1,\;2,\;3,\;4\,;\ny\\
\fl  Z_{ik}&=&\frac{Y_i}{Y_k}\,\hq\mbox{for}\hq ik=13,\;14,\;23,\;24;\hs 
  Z_{12}=Y_1Y_2;\hs Z_{34}=\frac{1}{Y_3Y_4}.\label{Y2}\eea 
and get
\beq \R_i^k=\lk\begin{array}{cc cc cc cc cc}
1 &\ds Z_{24}&0&0&0&0&\ds Z_{23} &\ds -Z_{34}\\[1mm]
\ds Z_{13}&\ds Z_{13}Z_{24}&0&0&0&0 &\ds-Z_{12} &\ds Z_{14}\\[1mm]
0&0& \ds Z_{13}Z_{24}& \ds Z_{13}& \ds Z_{14}&-Z_{12}&0&0\\[1mm]
0&0& \ds Z_{24}&1& \ds-Z_{34} &\ds Z_{23}&0&0\\[1mm]
0&0& \ds Z_{23} & \ds Z_{34}& 1 &\ds -Z_{24}&0&0\\[1mm]
0&0& \ds Z_{12} &\ds Z_{14} & \ds -Z_{13}&\ds Z_{13}Z_{24}&0&0\\[1mm]
\ds Z_{14}& Z_{12}&0&0&0&0&\ds Z_{13}Z_{24}&\ds -Z_{13}\\[1mm]
\ds Z_{34}&\ds Z_{23}&0&0&0&0&\ds -Z_{24}& 1\\
\end{array}\rk_{ik}\label{Vtm} \end{equation}
The determinant can be calculated directly: \beq 
\det\:\R\;=\;\lk Y_1\:Y_2\:\frac{(Y_3^2+1)(Y_4^2-1)}{Y_3^2\:Y_4^2}\rk^4\,.
\label{VTM}\end{equation}
For $N=2$ we can also write the MTE quite
explicitly. We shall write eq.(\ref{mte-elements}) shorthand as
\beq \Thop_{\:i_1,i_2,i_3,i_4,i_5,i_6}^{\:k_1,k_2,k_3,k_4,k_5,k_6}\;=\;
 \rho\;{\overline{\Thop}
 }_{\:i_1,i_2,i_3,i_4,i_5,i_6}^{\:k_1,k_2,k_3,k_4,k_5,k_6}.\label{mtnz}
 \end{equation}
where in obvious correspondence $\;\Thop\;$ and $\;\overline{\Thop}\,$
are defined to be the left- and right-hand sums of products of four 
$\:R$-matrices.
We use (\ref{Y2}) and abbreviate  $\;Y_i^{(j)}\;$ by $\;Y_{ij}\,,$
where $i$ labels the four points on the Fermat curve
$x_1,\;x_2,\;x_3,\;x_4=x_1x_2/(\om x_3)$, and $j=1,\ldots,8$ denote
the eight arguments of the $\Rop^{(j)}.$ The left hand side 
of $\:$(\ref{mtnz}) 
is, using $\:$(\ref{R-N-matrix}):
\bea\lefteqn{\Thop_{\:i_1,i_2,i_3,i_4,i_5,i_6}^{\:k_1,k_2,k_3,k_4,k_5,k_6}
  \;=\;(-1)^{k_3k_6}\;
  \frac{Y_{11}^{i_2+i_1}\;\:Y_{23}^{k_4+k_2}\;\:Y_{24}^{k_5+k_3}}
  {Y_{42}^{i_4+k_1}}\;\times}
  \ny\\&\times&\hspace*{-3mm}
\sum_{j_1,j_2,j_3,j_4,j_5,j_6}\hspace*{-2mm}
                   \delta_{i_2+i_3,j_2+j_3}\;\delta_{i_4+i_5,j_4+j_5}\;
       \delta_{j_4+i_6,k_4+j_6}\;\delta_{j_5+j_6,k_5+k_6}\;\times\ny\\
&\times&
\frac{(-1)^{j_1(j_3-j_5)+j_6(k_2-j_2)-j_3(i_1+k_6)+k_1j_5}
\;\;Y_{21}^{j_2+j_1}\;Y_{12}^{i_4+j_1}\;Y_{22}^{j_4+k_1}\;
Y_{13}^{j_4+j_2}\;Y_{14}^{j_5+j_3}}{Y_{31}^{j_2+i_1}\;Y_{41}^{i_2+j_1}
  \;Y_{32}^{j_4+j_1}\;Y_{33}^{k_4+j_2}\;Y_{43}^{j_4+k_2}\;Y_{34}^{k_5+j_3}
  \;Y_{44}^{j_5+k_3}},
\ny\eea
where all exponents are understood modulo 2, i.e. being just $0$ or $1$.
We can use three of the $\delta$'s to eliminate the sums over e.g.
$j_2,\;j_4,\;j_6$. Then the last $\delta$ gives a compatibility condition
with the result that if $\;i_4+i_5+i_6+k_4+k_5+k_6\;$ is odd, the component
of $\:\Thop\:$ vanishes and these components of the MTE are trivial. These
are half of the $\;2^{12}\;$ components. We also see that, if these 
occur at all,
$\;Y_{11},\;Y_{23},\;Y_{24}\:$ and $\:Y_{42}\:$
will factorize ($Y_{11}$ appears if $i_1+i_2$ is odd, etc.). Each non-zero 
component has a sum over eight terms on each side.
The result is:
\bea \lefteqn{\Thop_{\:i_1,i_2,i_3,i_4,i_5,i_6}^{\:k_1,k_2,k_3,k_4,k_5,k_6}
  \;=\;\delta_{i_4+i_5+i_6,k_4+k_5+k_6}(-1)^\gamma\;
  \frac{Y_{11}^{i_2+i_1}\;\:Y_{23}^{k_4+k_2}\;\:Y_{24}^{k_5+k_3}}
  {Y_{42}^{i_4+k_1}}\;\;\times  }\ny\\[2mm]&&
\sum_{j_1,j_2,j_4}\frac{(-1)^\beta\;\;Y_{21}^{j_2+j_1}
  \;Y_{12}^{i_4+j_1}\;Y_{22}^{j_4+k_1}\;
  Y_{13}^{j_4+j_2}\;Y_{14}^{j_4+j_2+\iota}}
  {Y_{31}^{j_2+i_1}\;Y_{41}^{i_2+j_1}
  \;Y_{32}^{j_4+j_1}\;Y_{33}^{k_4+j_2}\;
  Y_{43}^{j_4+k_2}\;Y_{34}^{k_5+i_2+i_3+j_2}
  \;Y_{44}^{i_4+i_5+j_4+k_3}}\label{Zwli}\eea  
where 
$\;\iota=i_2+i_3+i_4+i_5\;$ and 
\bea \beta&=& j_1j_2+j_2j_4+j_4j_1+j_1\,\iota
  +j_2(i_1+i_6+k_4+k_6)+j_4(k_1+k_2),      \ny\\ \fl\gamma&=&
  k_1(i_4+i_5)+k_2(i_6+k_4)+(i_2+i_3)(i_1+k_6)+k_3k_6.
  \ny\eea
The analogous expression for the right hand side is:
\bea \lefteqn{{\overline{\Thop}
 }_{\:i_1,i_2,i_3,i_4,i_5,i_6}^{\:k_1,k_2,k_3,k_4,k_5,k_6}\;
 =\;\delta_{i_4+i_5+i_6,k_4+k_5+k_6}(-1)^{\overline{\gamma}}\;
  \frac{Y_{25}^{k_2+k_1}\;\:Y_{18}^{i_5+i_3}\;\:Y_{17}^{i_4+i_2}}
  {Y_{36}^{k_4+i_1}}\;\;\times}\ny\\[2mm]&&
\sum_{j_1,j_2,j_4}
\frac{(-1)^{\overline{\beta}}\;\;Y_{16}^{j_4+i_1}\;Y_{15}^{j_2+j_1}
  \;Y_{26}^{j_1+k_4}\;
  Y_{27}^{j_4+j_2}\;Y_{28}^{j_4+j_2+\overline{\iota}}}{Y_{38}^{j_4+i_3+k_4+k_5}
  \;Y_{48}^{i_5+j_2+k_2+k_3}\;Y_{37}^{i_2+j_4}\;Y_{47}^{i_4+j_2}\;
  Y_{46}^{j_1+j_4}\;Y_{35}^{j_1+k_2}\;Y_{45}^{k_1+j_2}},
  \label{Zwri}\eea
with $\;\overline{\iota}=k_2+k_3+k_4+k_5,\;$ and
\bea \overline{\beta}&=&
j_1(k_3+k_5)+j_2(j_4+i_4)+j_4(k_2+k_3+i_3),  \ny\\
\fl \overline{\gamma}&=& i_1k_5+i_2k_6+k_1k_3+(i_4+k_6)(k_2+k_3+i_3).\ny\eea
and again all exponents are understood $ mod\;\:2$. 
\\[2mm]
We give two examples of non-trivial components of the MTEs:
First we consider the component 
\[\fl \Thop_{\;0,0,0,0,0,0}^{\;0,0,0,0,0,0}\;=\;
  \rho\;{\overline{\Thop}}_{\;0,0,0,0,0,0}^{\;0,0,0,0,0,0}\]
This is, written more explicitly, using the abbreviations of (\ref{Y2})  
adding the index $j$: $\;Z_{ik,j}\:=\:Y_i^{(j)}/Y_k^{(j)}$ 
for $ik=13,14,23,24\,,\;\;Z_{12,j}\:=\;Y_1^{(j)}Y_2^{(j)}\,,\;
Z_{34,j}\:=\:1/(Y_3^{(j)}\,Y_4^{(j)})$:
\bea 
\lefteqn{1+Z_{24,1}Z_{13,2}+(Z_{23,1}-Z_{34,1}Z_{13,2})Z_{13,3}Z_{13,4}}
\ny\\
\fl && +(Z_{23,2}-Z_{12,2}Z_{24,1})Z_{14,3}Z_{14,4}
 -(Z_{23,1}Z_{23,2}+Z_{34,1}Z_{12,2})Z_{34,3}Z_{34,4}\ny\\
\fl  &=&\rho\;\lb 1+Z_{24,6}Z_{13,5}+(Z_{14,5}+Z_{24,6}Z_{34,5})
Z_{24,8}Z_{24,7}
\right.\ny \\
\fl &&
\left.+(Z_{14,6}+Z_{12,6}Z_{13,5})Z_{23,8}Z_{23,7}-(Z_{14,6}Z_{14,5}
   +Z_{12,6}Z_{34,5})Z_{34,8}Z_{34,7}\rb. \label{exei}\eea
Another component:
\[\fl \Thop_{\;0,0,1,0,1,0}^{\;0,0,1,1,0,0}\;=\;\rho\:
 {\overline{\Thop}}_{\;0,0,1,0,1,0}^{\;0,0,1,1,0,0}\]
reads analogously:
\bea \lefteqn{(1+Z_{24,1}Z_{13,2})Z_{23,3}Z_{23,4}\;
 -(Z_{23,1}-Z_{13,2}Z_{34,1})
       Z_{12,3}Z_{12,4}}\ny\\
\fl  &&+(Z_{23,2}-Z_{24,1}Z_{12,2})Z_{13,3}Z_{24,3}Z_{13,4}Z_{24,4}\;
 +(Z_{23,1}Z_{23,2}+Z_{34,1}Z_{12,2})Z_{24,3}Z_{24,4}\ny\\
\fl &=&\rho\:\lb
 Z_{23,6}-Z_{34,6}Z_{13,5}\;
 +(Z_{23,6}Z_{14,5}-Z_{34,6}Z_{34,5})Z_{24,8}Z_{24,7}
 \right.\ny\\ \fl &&+\left. (Z_{24,6}-Z_{13,5})Z_{13,6}Z_{23,8}Z_{23,7}\;
 +(Z_{34,5}Z_{13,6}-Z_{13,6}Z_{24,6}Z_{14,5})Z_{34,7}Z_{34,8}\rb.
 \label{exzw}\eea
Each equation appears {\it eight} times for different components. 
In order to write the symmetries compactly,  
we introduce the following three mappings $\;a,\;b,\;c\;$ of the upper 
or lower indices: 
\bea  a(i_1,i_2,i_3,i_4,i_5,i_6)&=
          &(i_1+1,i_2+1,i_3,i_4+1,i_5,i_6);\ny\\  
\fl  b(i_1,i_2,i_3,i_4,i_5,i_6)&=
            &(i_1,\,i_2,\,i_3+1,\,i_4,i_5+1,\,i_6);\ny\\
\fl  c(i_1,i_2,i_3,i_4,i_5,i_6)&=
         &(i_1,\,i_2,\,i_3,\,i_4,\,i_5,\,i_6+1),\ny\\[2mm]
\fl \mbox{and}\hs\hs
 ab(i_1,i_2,i_3,i_4,i_5,i_6)&=&a(b(i_1,i_2,i_3,i_4,i_5,i_6))
         ,\hq etc.               
        \label{syla}  \eea
the same also for the $k_j$ instead of the $i_j$. Of course, 
the indices are always taken $\,mod\;\,2$. 
Altogether, for $N=2$ there are $2^8$ different nontrivial 
components, an independent set is (the same for the $\;{\overline\Thop}$)
\bea \lefteqn{\Thop_{i_1,i_2,i_3,0,0,0}^{k_1,k_2,k_3,0,0,0};\hs
\Thop_{i_1,i_2,i_3,0,0,0}^{k_1,k_2,k_3,1,1,0};\hs
\Thop_{i_1,i_2,i_3,0,0,0}^{k_1,k_2,k_3,1,0,1};\hs
\Thop_{i_1,i_2,i_3,0,0,0}^{k_1,k_2,k_3,0,1,1}\hspace{2cm}}\ny\\[2mm]
 \fl &&\mbox{for}\hspace{12mm} i_1,i_2,i_3,k_1,k_2,k_3=0,1.\eea
\begin{prop} For $N=2$ the components of the left-hand side of the 
Modified Tetrahedron Equation satisfy the following symmetry relations:
\bea &&\!\!\!\!\Thop_{i_1,i_2,i_3,0,0,0}^{k_1,k_2,k_3,k_4,k_5,k_6}\;=\;
  \Thop_{a(i_1,i_2,i_3,0,0,0)}^{a(k_1,k_2,k_3,k_4,k_5,k_6)}\ny\\
\fl &=&(-1)^{i_1+k_1}\Thop_{b(i_1,i_2,i_3,0,0,0)}^{b(k_1,k_2,k_3,k_4,k_5,k_6)}
      \;\;=\;
(-1)^{i_1+k_1}\Thop_{ab(i_1,i_2,i_3,0,0,0)}^{ab(k_1,k_2,k_3,k_4,k_5,k_6)}\ny\\
\fl &=&(-1)^{i_2+i_3+k_2+k_3}
  \Thop_{c(i_1,i_2,i_3,0,0,0)}^{c(k_1,k_2,k_3,k_4,k_5,k_6)}
       \;\;=\;(-1)^{i_2+i_3+k_2+k_3}
  \Thop_{ac(i_1,i_2,i_3,0,0,0)}^{ac(k_1,k_2,k_3,k_4,k_5,k_6)}\ny\\
\fl &=&(-1)^{i_1+\!i_2+\!i_3+\!k_1+\!k_2+\!k_3}
  \Thop_{bc(i_1,i_2,i_3,0,0,0)}^{bc(k_1,k_2,k_3,k_4,k_5,k_6)}=
(-1)^{i_1+\!i_2+\!i_3+\!k_1+\!k_2+\!k_3}
  \Thop_{abc(i_1,i_2,i_3,0,0,0)}^{abc(k_1,k_2,k_3,k_4,k_5,k_6)},
\ny\\
\fl &&  \label{symm} \eea  
where $\;i_1,i_2,i_3,k_1,k_2,\ldots, k_5,k_6=0,1.$
Here the mappings $a,\;b,\;c$ are defined as in (\ref{syla}). The same
equations are valid for the right-hand components, i.e. for $\Thop$ 
replaced by ${\overline\Thop}$. For $k_4+k_5+k_6\,$ odd,
these equations are trivial. \end{prop}
{\it Proof:}
Use the explicit formulae (\ref{Zwli}) and (\ref{Zwri}). For the relations
involving the mapping $a$ we shift the summation indices $j_1,\;j_2,\;j_4$.
Then all exponents of the $Y-$factors are unchanged. No phase is appearing, 
since the shift in $\gamma$ $({\overline\gamma})$ is compensated by the shift 
in $\beta$ $({\overline{\beta}})$. 
For the relations involving the mappings $b$ and/or $c$, also all exponents
of the $Y$ are unchanged. The phase factors appear from the shifts in $\gamma$
or ${\overline{\gamma}}$.  \hfill $\square$  
\\[3mm] 
The terms
\[\fl \hs\hq Z_{ik,j}\,=\,y_{ik}^{(j)}\;\frac{1+x_k^{(j)}}{1+x_i^{(j)}}\,, 
\hs ik=13,\;14,\;23,\;24;\] \vspace{-4mm} 
\[\fl Z_{12,j}=\frac{y_{13}^{(j)}\,y_{23}^{(j)}\,(1-{x_3^{(j)}}^2)}
{(1+x_1^{(j)})(1+x_2^{(j)})}\,;\hs\hx 
Z_{34,j}=\frac{(1+x_3^{(j)})(1+x_4^{(j)})}{y_{31}^{(j)}\,y_{41}^{(j)}
\,(1-{x_1^{(j)}}^2)}\,,\hs\]
which appear in (\ref{exei}) and (\ref{exzw}) have the following explicit
form, taking all boundary coefficients and all $\ka_j$ 
to be unity, compare (\ref{resca}),(\ref{rescb})\,:
\bea Z_{13,1}&=& \frac{({b_2}'c_2-b_1)d_2}{c_2-ib_2};\hs
 \hx\; Z_{14,1}\;=\;
 \frac{(b_2c_2'-c_1)d_2}{(b_2+ic_2)d_1};\ny\\
\fl Z_{23,1}&=&\frac{({b_2}'c_2-b_1){d_2}'}{c_1{b_2}'-ib_1c_2'};\hs\hx\;
Z_{24,1}\;=\;\frac{(b_2c_2'-c_1){d_2}'}{(b_1c_2'+i{b_2}'c_1)d_1};\ny\\[1mm]
\fl  &\ldots&\ny\\ 
\fl Z_{23,4}&=&\frac{(a_2^{\dag\dag}{b_2}'-a_1^\dgh)c_2^\dgh}
           {{b_1}'''a_2^{\dag\dag}-ia_1^\dgh{b_2}^{\dag\dag}};\hs\!\!
Z_{24,4}\;=\;\frac{(a_2''b_2^{\dag\dag}-b_1''')c_2^\dgh}
  {(a_1^\dgh b_2^{\dag\dag}+ia_2^{\dag\dag}b_1''')c_1''};
 \label{zin}    \eea
etc. Discrete sign choices of square roots come in when expressing the 
transformed variables e.g. $\;b_2',\;a_1''',$ etc. in terms of the original 
eight variables $a_1,\;a_2,\;\ldots,\;d_2\;$ via (\ref{deri}). E.g. from
the first lines of (\ref{deri}):
\[\fl b_2'\,=\,\pm\frac{1}{c_2d_2}\sqrt{b_1^2d_2^2+ b_2^2+c_2^2}\:;\hs
 c_2'\,=\,\pm\frac{1}{b_2d_2}\sqrt{c_1^2d_2^2+(b_2^2+c_2^2)d_1^2}\,;\]
\[\fl {d_2}'\!=\,\pm\frac{1}{b_2c_2}
         \sqrt{(b_2^2+b_1^2d_2^2)c_1^2+b_1^2c_2^2d_1^2}\:;\hs\!\!
a_2''\,=\,\pm\frac{1}{c_1d_1}
\sqrt{(a_1^2d_1^2+a_2^2{d_2}'^2){c_2'}^2+c_1^2{d_2}'^2}\,.\]
All dashed or daggered variables are square roots of rational expressions.
\\[2mm]
From (\ref{VTM}) the factor $\rho$ is:
\beq \rho\;=\;\sqrt{\frac{\prod_{i=1}^4\;Z_{12,i}
\lk 1+Z_{34,i}(Z_{13,i}Z_{14,i}^{-1}-Z_{14,i}Z_{13,i}^{-1})-Z_{34,i}^2\rk}
{\prod_{j=5}^8\;Z_{12,j}
\lk 1+Z_{34,j}(Z_{13,j}Z_{14,j}^{-1}-Z_{14,j}Z_{13,j}^{-1})
    -Z_{34,j}^2\rk}}\:. \label{ro}\end{equation}
Using the components like (\ref{exzw}) and inserting there (\ref{ro}) and 
(\ref{zin}),~(\ref{deri}) we get the MTE in terms of our eight parameters
$a_1,\;\ldots,\:d_2$ and sixteen choices of the signs of
\beq  {b_2}';\;c_2';\;{d_2}';\;a_2'';\;c_1'';\;d_1'';\;b_1''';\;d_2''';
\hx         
  a_1^\dg;\;b_1^\dg;\;c_1^\dg;\;a_2^\dgg;\;b_2^\dgg;\;d_1^\dgg;\;
   a_1^\dgh;\;c_2^\dgh.\label{sgnc}\end{equation}     
We have confirmed the $N=2$-MTE numerically for all 256 different components, 
choosing random complex numbers for the eight continuous variables and random 
signs for the sixteen square roots in the variables (\ref{sgnc}).

We conclude mentioning that for $N=3$ we find that of the $3^{12}$ components
of the MTE for 
$\;\lk R^{(j)}\rk_{i_1,i_2,i_3}^{j_1,j_2,j_3}\;$ given in (\ref{R-N-matrix}),
$\,2 \times 3^{11}$ components are just $0\;=\;0,\;$
while $3^{11}$ (not all distinct) equations have non-trivial 
left- and right-hand sides.

\section{Conclusions}
In this paper we study the Modified Tetrahedron Equation (\ref{mte-elements}) 
in which the Boltzmann weights $\;R_{i_1,i_2,i_3}^{j_1,j_2,j_3}\;$ depend on 
Fermat-curve variables via cyclic weight functions, see (\ref{R-N-matrix}). 
The conjugation by the $R$-matrix is a rational automorphism of the 
ultra-local 
Weyl algebra at the $N$-$th$ root of unity. The representation of this 
automorphism as a functional mapping in the space of the parameters of the 
ultra-local Weyl algebra allows us to obtain the free parameterization of 
the MTE. 
By "free parameterization" we mean that we leave free which solution of the
functional tetrahedron equation (\ref{fc-te}) with which boundary conditions 
will be chosen. 
We express the Fermat-curve variables (and so the Boltzmann weights) in terms 
of an independent set of eight continuous parameters and specify the sixteen 
phases which can be chosen independently. We derive a general expression for
the scalar factor of the MTE.  
For the simplest non-trivial case $N=2$ the MTE is written out explicitly. 
In this case it contains 256 linearly independent components.

The MTE allow to obtain a wide class of new integrable models: 
The $\R$-matrices may be combined into the cubic blocks obeying
globally the usual TE by due to the validity of the local MTE-s. 
The advantage of 
the free parameterization presented here is that it allows to get the 
appropriate parameterization for blocks of any size. This is the subject of a
forthcoming paper.

New integrable 2-dimensional lattice models with parameters living on
higher Riemann surfaces can be obtained from the MTE by
a contraction process which has been described in \cite{s-rma}. 

A further important application of the MTE concerns the following:
As usual the TE leads to the commutativity of the layer-to-layer transfer 
matrices, while the MTE can be used to obtain \emph{exchange relations} for 
the layer-to-layer transfer matrices. The exchange relations are related to 
isospectrality deformations and form the basis for a functional Bethe ansatz 
for three dimensional integrable spin models, see \cite{svan}.

Note finally, that starting from the results of this paper, we can consider 
several limits of the parameterization, which are connected to various 
degenerations of the weights. 
The usual Tetrahedron equation of \cite{sms-vertex} follows from the MTE in 
the special regime when the free parameters $u_j,w_j$, $j=1\ldots 6$ belong 
to the submanifold of $\mathbb{C}^{12}$ defined by
\begin{equation}\label{fid}\fl
\ds u_l^N-\Rop^{(f)}_{i,j,k}\cdot u_l^N\;=\;
w_l^N-\Rop^{(f)}_{i,j,k}\cdot w_l^N\;=\;0\;.
\end{equation}
This variety may be parameterized in terms of spherical geometry data. E.g. 
$\R_{1,2,3}$ may be associated with the spherical triangle with the
dihedral angles $\theta_1,\theta_2,\theta_3$ and
\begin{equation}\fl
\kappa_1^N\;=\;\tan^2\frac{\theta_1}{2}\;,\;\;\;
\kappa_2^N\;=\;\cot^2\frac{\theta_2}{2}\;,\;\;\;
\kappa_3^N\;=\;\tan^2\frac{\theta_3}{2}\;.
\end{equation}

\ack{
This work was supported in part by the contract INTAS OPEN
00-00055 and by the Heisenberg-Landau program HLP-2002-11.
S.P.'s work was supported in part by the grants 
CRDF RM1-2334-MO-02, RFBR 01-01-00539 and the 
grant for support of scientific schools
RFBR 00-15-9655. S.S.'s work was supported in part by the 
grants CRDF RM1-2334-MO-02 and RFBR 01-01-00201.}

\section*{References}
\bibliographystyle{amsplain}

\begin{thebibliography}{**}
\bibitem{Zamo} Zamolodchikov A B 1981 
\emph{Commun. Math. Phys.} {\bf 79} p~489
\bibitem{Bazh-Bax} Bazhanov V V and Baxter R J
1992 \emph{J. Stat. Phys.}
{\bf 69} p~453; 1993 {\bf 71} p~839
\bibitem{bms-mte} Boos H E, Mangazeev V V and Sergeev S M
1995  \emph{Int. J. Mod. Phys. A} {\bf 10} p~4041
\bibitem{sms-vertex} Sergeev S M, Mangazeev V V and Stroganov Yu G
1996 \emph{J. Stat. Phys.} {\bf 82} p~31
\bibitem{s-old} Sergeev S M, Bazhanov V V and Mangazeev V V 1995
\emph{Quantum Dilogarithm and Tetrahedron Equation} Preprint IHEP 
95 -- 129 ({\it unpublished}),\\ 
Sergeev S 1996 \emph{Operator solutions of simplex equations}
In {\it Proceedings of the 10th International Conference on
Problems of Quantum Field Theory}, ed. by Shirkov D V, 
Kazakov D I, Vladimirov A A, published by 
Publishing Department JINR, 1996, ISBN 5-85165-455-4
p~154
\bibitem{ms-mte} Maillard J-M and Sergeev S M
1997 \emph{Phys. Lett.} {\bf B405} p~55
\bibitem{s-kirhgoff} Sergeev S M 1997 
\emph{On a two dimensional system associated with the complex of the
solutions of  the Tetrahedron equation}
solv-int/9709013
\bibitem{s-symplectic} Sergeev S M 1998
\emph{Phys. Lett.} {\bf A 253} p~145
\bibitem{s-qem} Sergeev S M 1999
\emph{J. Phys. A: Math. Gen.} {\bf 32} p~5639
\bibitem{s-kiev} Sergeev S M 2001
\emph{Integrable three dimensional models in wholly discrete
space-time}
in {\it Integrable Structures of Exactly Solvable
Two-Dimensional Models etc.}, eds. Pakuliak S and
von Gehlen G, NATO Science Series II {\bf 35}, Kluwer, Dordrecht (2001), 
p. 293-304
\bibitem{Fa-Vo} Faddeev L D and Volkov A Yu 1992 
\emph{Theoret. and Math. Phys.}
{\bf 92} p~837
\bibitem{fk-qd}
Faddeev L D and Kashaev R M 1994
\emph{Mod. Phys. Lett.} {\bf A9} p~427
\bibitem{br-qd}
Bazhanov V V and Reshetikhin N Yu 1995
\emph{J. Phys. A: Math. Gen.} {\bf 28} p~2217
\bibitem{br-unpublished}
Bazhanov V V and N. Yu. Reshetikhin N Yu 1995 \emph{Chiral Potts model and
discrete Sine-Gordon model at roots of unity} unpublished
\bibitem{s-rma} Sergeev S M 1995 \emph{Mod. Phys. Lett.} {\bf A12} p~1393
%
\bibitem{s-tmf1} Sergeev S M 1999
\emph{Theoret. and Math. Phys.} {\bf 118} p~378, 
2000 {\bf 124} p~1187 
\bibitem{Bazh-Bax-st-tri} Bazhanov V V and Baxter R J 1993
\emph{J. Stat. Phys.} {\bf 71} p~839
\bibitem{svan}
Sergeev S M 2002 \emph{Functional equations and 
quantum separation of variables
for 3D spin models} Preprint MPI 02-46.
\end{thebibliography}

\end{document}